\documentclass[times,twocolumn,final]{elsarticle}
\usepackage{medima}
\usepackage{framed,multirow}
\usepackage{amssymb}
\usepackage{latexsym}
\usepackage{amsmath}
\usepackage{url}
\usepackage{xcolor}
\usepackage{diagbox}
\usepackage{algorithm}
\usepackage{algorithmic}
\usepackage{tabularx} 
\usepackage{adjustbox}
\usepackage{multirow}
\usepackage{makecell}
\usepackage{booktabs}
\usepackage{xcolor} 
\usepackage[table]{xcolor}
\usepackage{subcaption}

\usepackage{float}
\usepackage{subcaption}
\usepackage{xcolor} 
\usepackage{ulem}
\usepackage{hyperref}
\usepackage{color}
\usepackage{hyperref}
\usepackage{multirow}
\usepackage{amssymb}
\usepackage{booktabs}
\usepackage{xcolor} 
\usepackage[table]{xcolor}
 
\definecolor{best}{RGB}{188,188,223}
\definecolor{second}{RGB}{218,208,231}
\definecolor{third}{RGB}{234,237,247}
\usepackage[switch]{lineno}

\journal{Medical Image Analysis}

\begin{document}

\verso{Given-name Surname \textit{et~al.}}

\begin{frontmatter}

\title{Generative Data-engine Foundation Model for Universal Few-shot 2D Vascular Image Segmentation}

\author[ins4]{Rongjun Ge\fnref{equcont}}
\author[ins1]{Xin Li\fnref{equcont}}
\author[ins1]{Yuxing Liu}
\author[ins6]{Chengliang Liu}
\author[ins1]{Pinzheng Zhang}
\author[ins5]{Jiong Zhang}
\author[ins8]{Jian Yang}
\author[ins7]{Jean-Louis Dillenseger}
\author[ins1]{Chunfeng Yang\corref{mycorrespondingauthor}}\ead{chunfeng.yang@seu.edu.cn}
\author[ins3]{Yuting He\corref{mycorrespondingauthor}}\ead{yuting.he4@case.edu}
\author[ins1]{Yang Chen}

\cortext[mycorrespondingauthor]{Corresponding author}
\fntext[co-first]{Equal contribution}

\address[ins4]{School of Instrument Science and Engineering, Southeast University, Nanjing, China}
\address[ins1]{Key Laboratory of New Generation Artificial Intelligence Technology and Its Interdisciplinary Applications (Southeast University), Ministry of Education, Nanjing, China}
\address[ins6]{Department of Computer and Information Science, University of Macau, Macau, China}
\address[ins5]{Laboratory of Advanced Theranostic Materials and Technology, Ningbo Institute of Materials Technology and Engineering, Chinese Academy of Sciences, Ningbo, China}
\address[ins8]{Beijing Engineering Research Center of Mixed Reality and Advanced Display, School of Optics and Photonics, Beijing Institute of Technology, Beijing, China}
\address[ins7]{National Institute of Health and Medical Research, 35000 Rennes, France}
\address[ins3]{Department of Biomedical Engineering, Case Western Reserve University, Cleveland, USA}
\received{1 May 2013}
\finalform{10 May 2013}
\accepted{13 May 2013}
\availableonline{15 May 2013}
\communicated{S. Sarkar}

\begin{abstract} 
The segmentation of 2D vascular structures via deep learning holds significant clinical value but is hindered by the scarcity of annotated data, severely limiting its widespread application. Developing a universal few-shot vascular segmentation model is highly desirable, yet remains challenging due to the need for extensive training and the inherent complexities of vascular imaging. In this work, we propose \textbf{UniVG} (Generative Data-engine Foundation Model for Universal Few-shot 2D Vascular Image Segmentation), a novel approach that learns the compositionality of vascular images and constructing a generative foundation model for robust vascular segmentation. UniVG enables the synthesis and learning of diverse and realistic vascular images through two key innovations: \textit{1) Compositional learning} for flexible and diverse vascular synthesis: It decomposes and recombines vascular structures with varying morphological features and diverse foreground-background configurations to generate richly diverse synthetic image-label pairs. \textit{2) Few-shot generative adaptation} for transferable segmentation: It fine-tunes pre-trained models with minimal annotated data to bridge the gap between synthetic and real vascular domains, synthesizing authentic and diverse vessel images for downstream few-shot vascular segmentation learning. To support our approach, we develop UniVG-58K, a large dataset comprising 58,689 vascular images across five imaging modalities, facilitating robust large-scale generative pre-training. Extensive experiments on 11 vessel segmentation tasks cross 5 modalties (only with 5 labeled images on each task) demonstrate that UniVG achieves performance comparable to fully supervised models, significantly reducing data collection and annotation costs. All code and datasets will be made publicly available at \url{https://github.com/XinAloha/UniVG}.

\end{abstract}


\end{frontmatter}

\section{Introduction}
\label{sec:introduction}

The segmentation of 2D blood vascular holds significant clinical value in disease diagnosis, screening, and surgical planning and navigation \citep{Moccia2018Blood}. However, its application has been significantly impeded due to the inability to construct large amounts of 2D datasets for training, as two factors contribute to this limitation: \textbf{1)} Scarcity of vascular images stem from high collection costs, privacy protection policies, and the complex processes of data sharing among hospitals \citep{dumont2021overcoming}. These factors make it particularly difficult to collect sufficiently large and diverse datasets from clinical settings, thereby limiting the generalization and performance improvement of models.  \textbf{2)} Shortage of high-quality vascular annotations results from the requirement for specialized doctors to dedicate extensive time to performing pixel-level annotations on images \citep{khan2019review}. The elongated nature of vascular structures makes the annotation process even more time-consuming \citep{yagis2024deep}, thereby limiting the construction of large amounts of labeled datasets. Unlike conventional organ segmentation where anatomical structures maintain relatively consistent morphologies, vascular networks exhibit rich structural diversity with complex branching patterns and highly variable topological structures, which significantly increases the complexity of feature representation and annotation requirements \citep{zhang2025pasc}. A promising solution involves training deep learning models to achieve precise segmentation with few samples, thereby significantly reducing data acquisition costs while streamlining the model development pipeline \citep{song2023comprehensive}.

\begin{figure}[t!]
    \centering    
    \includegraphics[width=0.5\textwidth,height=0.6\textwidth]{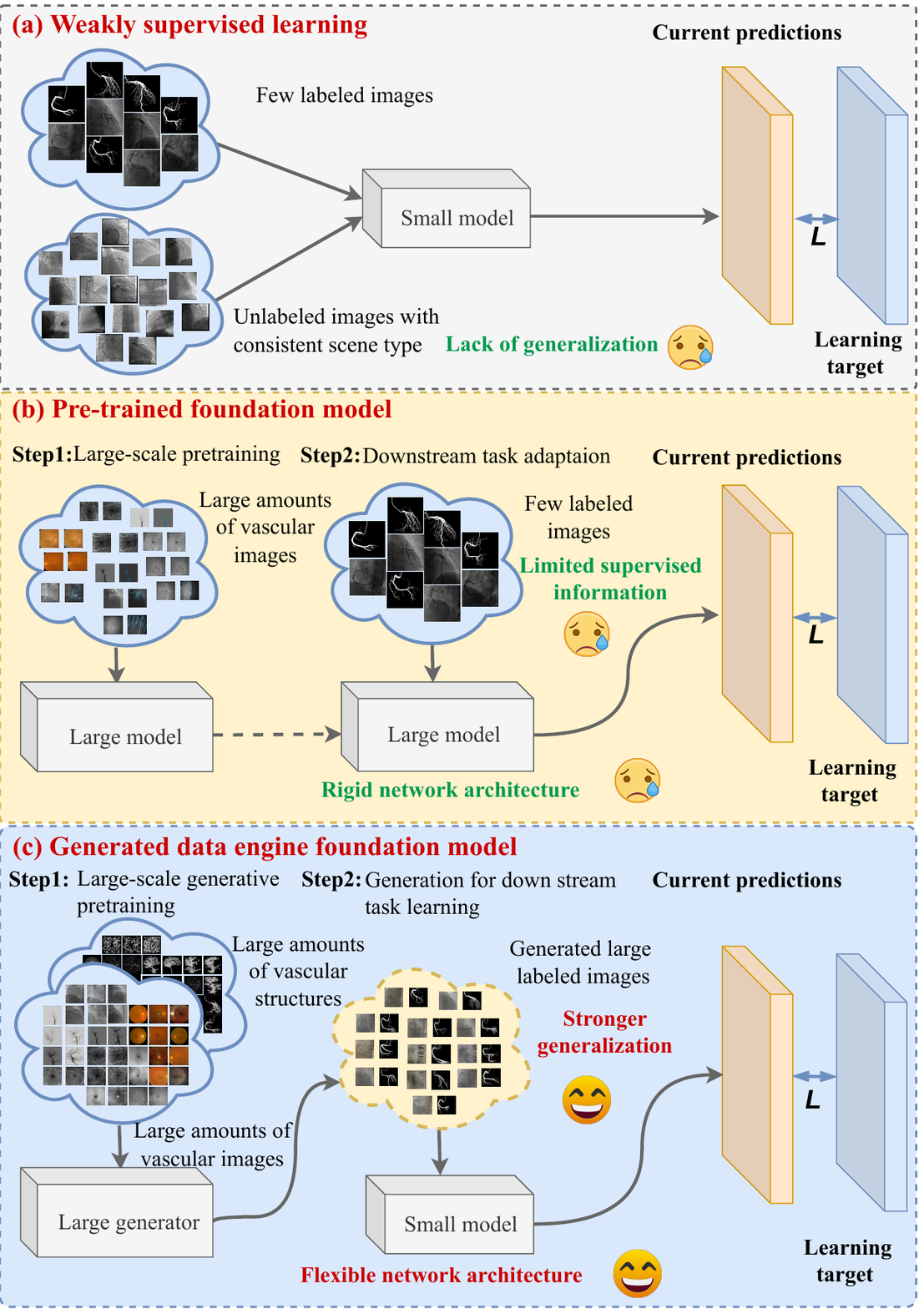}
    \caption{The advantages of our generative data-engine foundation models are as follows, (a) Weakly supervised learning: It utilizes a small number of sample images and unlabeled but consistent scene types. This model exhibits limited generalization capability. (b) Foundation models developed via large amounts of images pretraining adopt fixed architectural structures. When adapted to downstream tasks with few annotated images, these models encounter dual constraints: insufficient supervisory signals and inflexible network architectures. (c) In our generative data-engine foundation model, the first phase conducts pretraining of the generative model by integrating both collected large amount images and structured data synthesized through rule-based vascular generation algorithms. During the second phase of downstream adaptation, the model not only generates images at scale but also dynamically selects optimal architectures tailored to task-specific requirements.}
    \label{fig:advantage}
\end{figure}

Some studies  \citep{kozinski2020tracing,dang2022vessel, wu2023sparsely, chen2021semi, ouali2020semi}  attempt to use Weakly supervised learning (WSL) to reduce the cost of 2D vascular annotation by leveraging a small number of labeled 2D images and a large amount of unlabeled data \citep{zhou2018brief}. However, this approach is limited by the volume of available data and the quality of weak supervision information \citep{lin2023yolocurvseg}. As shown in Fig. \ref{fig:advantage} (a), WSL reduces the reliance on extensive labeled data by leveraging a small number of annotated images and unlabeled data with consistent scene types for model training, its application faces several challenges. On one hand, models trained using such methods are typically task-specific and struggle to generalize to other vascular scenarios \citep{shu2022cross}. On the other hand, the quality of weak supervision information directly impacts model performance \citep{lin2016scribblesup}, as inaccurate or incomplete labels lead to degraded outcomes, particularly when handling complex structures such as vascular \citep{tang2018normalized,kervadec2019constrained}.

In recent years, Pre-trained Foundation Models (PFM) have garnered significant attention across various domains \citep{he2024foundation,bommasani2021opportunities}. As illustrated in Fig. \ref{fig:advantage} (b), these models establish effective prior knowledge by learning generic feature representations through large-scale pretraining , and demonstrate robust transfer learning capabilities during downstream task adaptation using few labeled images. This approach substantially reduces reliance on annotated target domain data and enhances model convergence efficiency. However, in the field of 2D vascular image analysis, the performance of such models faces several critical challenges, primarily including domain discrepancy, limited supervised information, and rigid network architecture. Specifically, 1) Domain discrepancy: Vascular images exhibit significant differences from natural images in terms of morphological characteristics, contrast distribution, and topological structures \citep{raghu2019transfusion,HU2024103164}. This results in a significant mismatch between the pre-trained feature space and the feature space required for the vessel segmentation task, thereby leading to a degradation in model performance.
2) Limited supervised information: 
When fine-tuning with few labeled image datasets, the model is prone to overfitting, particularly in high-dimensional parameter spaces with insufficient supervisory signals \citep{tajbakhsh2020embracing}. This limitation further constrains the model's generalization capability. 3) Rigid network architecture: Existing pre-trained models employ fixed network architectures, hindering task-specific optimization due to their inherent structural constraints \citep{ma2022rethinking, isensee2021nnu} . This architectural rigidity limits the model's capacity to capture intricate vascular features effectively. This constraint particularly affects vascular segmentation, as specialized networks designed for elongated structures with multi-scale feature extraction capabilities cannot be utilized within  pre-training transfer paradigms \citep{yang2024enhancing}.

Existing research indicates that the learning mechanisms of neural networks possess the capability to extract concepts from a limited number of samples \citep{Sorscher2022Neural}, a capability partly attributed to their ability to decompose and recompose entities, known as compositionality \citep{mishra2022interpretable}. Compositionality refers to the process of decomposing and recombining entities to generate new samples, thereby enabling the model to adaptively understand the underlying logic of the entities. As illustrated in Fig. \ref{fig:combination} (a), the branch structures of plants can be decomposed and recombined to generate plant morphologies with distinct branching characteristics, further combining plants with pots can create entirely new plant-pot composite samples. This process of generating new samples through composition not only enhances data diversity but also allows the learning mechanism to gain a deeper understanding of the inherent patterns and relationships underlying the entities. Based on this, we propose a hypothesis: by decomposing and recombining vascular images, integrating compositionality into the learning process can enable the development of a universal few-shot vascular segmentation model. This approach aims to leverage compositionality to generate diverse vascular image samples, thereby enhancing the model's generalization capability and segmentation accuracy few annotated data.

In this study, we integrate compositional modeling into few-shot vascular segmentation learning, constructing a generative-based universal framework for vascular segmentation. As illustrated in Fig. \ref{fig:combination} (b), the core concept involves decomposing and recombining vascular structures, analogous to compositional process observed in plants, and combining these structures with diverse backgrounds to generate novel vascular images. Building on this innovative approach, we develop the compositional learning for flexible and diverse vascular synthesis. As depicted in Fig. \ref{fig:advantage} (c), this method uses rich vascular structures and vascular images for conducting large-scale generative pretraining. To mitigate the domain discrepancy between generated data and target segmentation images, we proposed a few-shot generative adaptation for transferable vascular image segmentation. Subsequently, the trained synthesizer generates a rich set of labeled images to train the segmentation model.  The generated data is highly diverse and rich, which substantially enhances the model’s generalization capabilities. Furthermore, the downstream segmentation network can be flexibly replaced with alternative architectures, improving the framework’s universality and adaptability. Building upon the aforementioned methodological innovations,  we ultimately establish the  Generative Data-engine Foundation Model for Universal Few-shot Vascular Image Segmentation (UniVG). By embedding compositionality into the vascular learning process, our method achieves performance comparable to or exceeding fully supervised models using five annotated images in arbitrary scenarios, offering an efficient solution for few-shot vascular segmentation learning. Overall, our proposed UniVG framework encompasses two innovations:

\begin{figure}
    \raggedright    \includegraphics[width=0.5\textwidth]{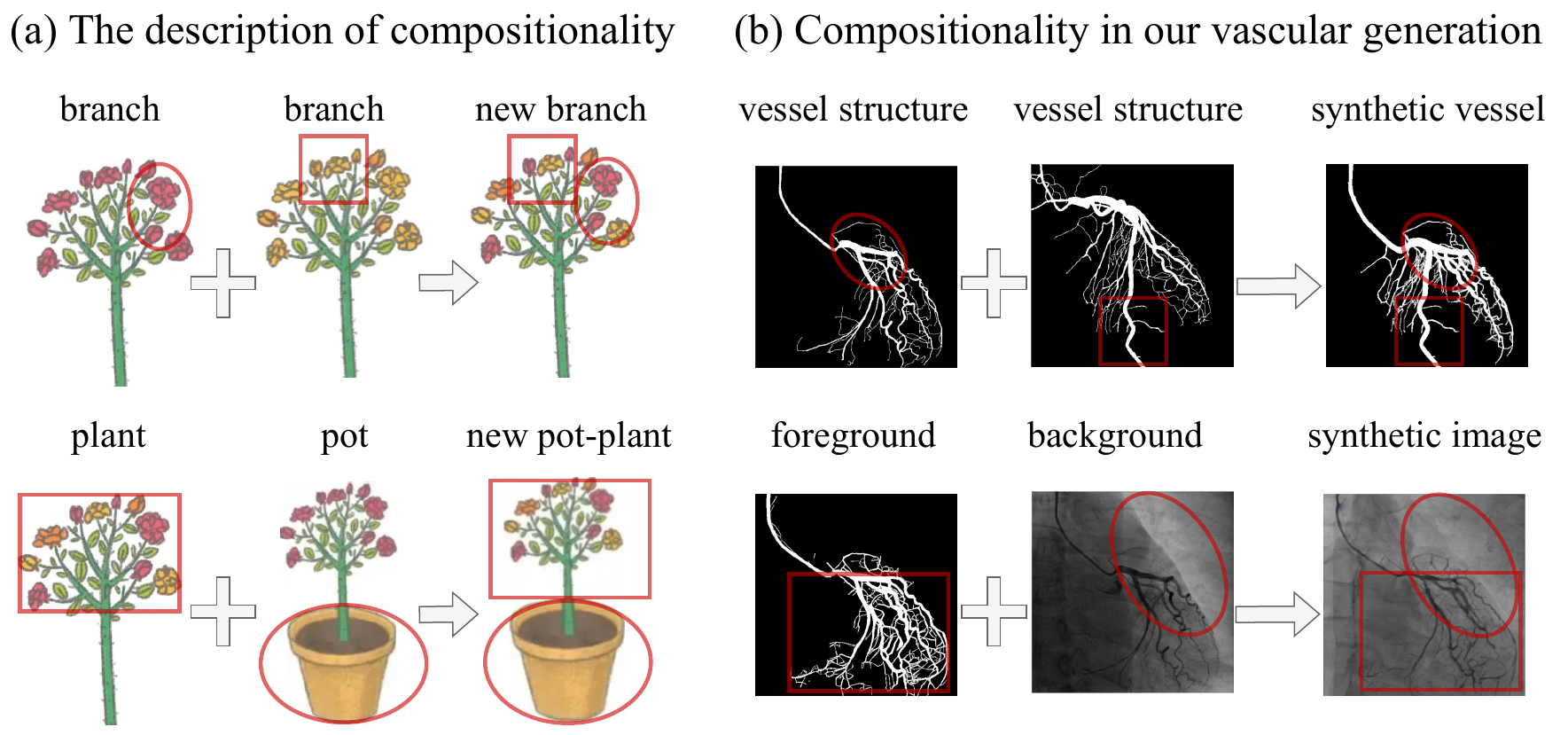}
    \caption{The motivation in our generative data-engine foundation model is that compositionality can enhance the diversity of images. a) The description of compositionality in plants, two branches combined into a new branch and a plant
    and a pot combined into a new pot-plan.b) Compositionality in our vascular generation is reflected in the synthesis of new vascular structures by combining different vascular structures, as well as the combination of background and foreground to create new vascular images.}
    \label{fig:combination}
\end{figure}

\noindent
1) \textit{Compositional learning for flexible and diverse vascular synthesis}: Through compositional learning, the model can learn the structural features of vascular structures at various levels and the distribution characteristics of vascular images, thereby enhancing the diversity of synthesized images. Specifically, this mainly includes the following two aspects:

a) Enhancing the diversity of 2D vascular structures: Our approach applies rule-based vascular generation algorithms to flexibly produce pseudo-vascular structures with various morphological features. This method allows for customization of the complexity and layout of vascular structures by adjusting parameters and configurations during the generation process, thereby naturally generating a diversified range of vascular structures. Subsequently, a diffusion model is utilized for pre-training to learn representations of these structures at different types and levels, integrating multi-level knowledge to generate an even broader array of vascular structure combinations. This not only supports the creation of a richly diverse set of vascular structures but is also particularly suitable for vascular segmentation tasks with few samples.

b) Increasing the diversity of 2D vascular image styles: To further enrich the styles of vascular structures, we constructed a dataset named UniVG-58K, comprising 57,075 unlabeled vascular images. This dataset was used to pre-train the diffusion model, enabling it to capture the rich characteristics of the distribution of vascular images. Ultimately, by integrating vascular structures with background information and diverse vascular masks, our method generates stylistically varied vascular images, improving dataset diversity and practicality.

\noindent
2) \textit{Few-shot generative adaptation for transferable vascular image segmentation}: In the process of adaptation via fine-tuning, embedding few-shot real data enables generative models to produce high-quality synthetic data. Such data are leveraged by downstream networks to learn effective domain-specific information, ultimately achieving performance that approaches or even surpasses fully supervised models trained on real datasets.

a) Few-shot transfer of 2D vascular structure knowledge: By fine-tuning a generative model that has learned from a rich combination of vascular structures with a small amount of real vascular masks, we aim to transfer foundational knowledge of vascular structures into corresponding real-world vascular contexts. Consequently, this approach facilitates the generation of  vascular structures that are both diverse and realistic, significantly advancing further research in vascular generation. 

b) Few-shot transfer of 2D vascular image knowledge: Through the use of a few annotated vascular images for transfer learning on a conditional generative model that has been pretrained on a large ensemble of vascular image combinations, we conditionally generate corresponding  vascular images based on the vascular structures produced as described in a). This adapted model generates vascular images with diverse structures and styles, emulating the complexity of real vascular distributions. Ultimately, the integration of a large volume of synthetic images alongside a few real data trains downstream segmentation networks to extract domain-invariant features, enabling robust generalization in unseen vascular domains.

In summary, the contributions of this paper are as follows:
\begin{itemize}
    \item To the best of our knowledge, we  propose a novel generative data-engine foundation model for universal few-shot vascular image segmentation for the first time. This innovative approach enables the generation of sufficiently diverse vascular images even under few-shot conditions, thereby providing essential data support for downstream vascular segmentation tasks. Remarkably, it achieves performance comparable to fully supervised models trained on real datasets using as few as five annotated vascular images in any 2D vascular segmentation scenario.
    
    \item We propose compositional learning for flexible and diverse vascular synthesis, where different branches of vascular structures are recombined and integrated with diverse foreground-background environments, achieving flexible and highly diversified labeled vascular image generation. 
    
    \item We propose a few-shot generative adaptation for transferable vascular image segmentation, specializing knowledge from pre-trained generation models for the real vascular domain, enabling the generation of high-quality synthetic data from a few annotated data. This supports effective training and generalization of downstream segmentation models.

    \item We develop a large-scale vascular image dataset, UniVG-58K, containing 57,075 vascular images, and used vascular generation algorithms to create numerous pseudo-vascular structures. Based on this dataset, large-scale pre-training was conducted, significantly enhancing the structural and stylistic diversity of vascular images.
    
    \item We conducted extensive experiments on 11 vascular segmentation tasks cross 5 modalities to verify the effectiveness of our proposed UniVG framework. The results demonstrate that our method can generate diverse vascular images and effectively enhance the performance of segmentation models in 5-shot scenarios. This confirms the robustness and versatility of our framework across various applications.
\end{itemize}

Overall, our UniVG framework has three key advantages: 1) \textbf{Diversity}: Our UniVG  framework can generate morphologically diverse vascular images using only a small amount of annotated data, effectively augmenting the data for few-shot vascular segmentation tasks. This significantly reduces the labeling costs in vascular image segmentation tasks.
2) \textbf{Transferability}: Our UniVG framework demonstrates excellent versatility, requiring only 5 images and corresponding masks to be applied to different datasets and segmentation tasks.
3) \textbf{Accuracy}: Compared to other segmentation methods, the authentic images generated by our framework significantly improve the performance of the segmentation model. Our UniVG framework achieves state-of-the-art results in multiple vascular segmentation tasks. We implement our proposed method using the PyTorch framework and share our code at \url{https://github.com/XinAloha/UniVG}. 

\section{Related Works}
\label{sec:related work}
\subsection{Pre-trained foundation model}
Pre-training is a critical phase of foundation models, significantly enhancing model performance through large-scale data learning. Compared to models trained from scratch, pre-trained models exhibit higher robustness and better generalization capabilities across multiple tasks \citep{yang2024vilref}. In the medical domain, the application of foundation models has been shown to significantly improve diagnostic accuracy, accelerate disease screening, assist clinical decision-making, reduce healthcare costs, and advance medical research, thereby comprehensively enhancing the quality and efficiency of medical services \citep{Zhang2023On}.

Foundation models have demonstrated significant advantages in the field of medical image segmentation. For instance, BiomedParse \citep{zhao2025foundation} was proposed as a biomedical foundation model capable of jointly executing segmentation, detection, and recognition tasks across up to nine imaging modalities. This joint task learning not only improves the accuracy of individual tasks but also opens new applications such as segmenting relevant objects in images through textual descriptions. Additionally, \citet{zhang2025foundation} explored a foundation model for brain MRI lesion segmentation, utilizing a mixture of modality experts framework to automatically segment different types of brain lesions, exhibiting excellent generalization capability across multiple input modalities. \citet{berger2025cross} introduced cross-domain transfer learning for image-to-graph transformers, enabling 2D road network data to pretrain 3D vessel graph extraction models through regularized edge sampling and supervised domain adaptation frameworks. \citet{wittmann2025vesselfm} proposed vesselFM for universal 3D blood vessel segmentation, trained on three heterogeneous data sources, achieving zero-shot generalization across diverse imaging modalities such as MRA, CTA, vEM, and microscopy images. 

The pre-training strategies for foundation models can be primarily categorized into two types: supervised pre-training and self-supervised pre-training. Supervised pre-training leverages labeled data to train the model, enabling it to learn the mapping relationship between input data and labels, which improves the model's accuracy and generalization capabilities. For instance, SAM and MedSAM \citep{ma2024segment} have demonstrated outstanding performance in natural image and medical image segmentation tasks DeepVesselNet \citep{tetteh2020deepvesselnet} alleviates the problem of insufficient annotated data by generating synthetic 3D vascular datasets and applying transfer learning. Self-supervised pre-training, on the other hand, utilizes unlabeled data to learn the intrinsic structures and patterns within the data, mapping the input data to a low-dimensional representation to better capture the essential features. Self-supervised pre-training methods, such as MoCo \citep{he2020momentum} , MAE \citep{he2022masked}, SimCLR \citep{liu2022deep}, iBOT \citep{zhou2021ibot}, and SimSiam \citep{chen2021exploring} , have achieved significant results in various tasks. These methods enhance the model's representation capabilities by minimizing reconstruction objectives or maximizing similarity measures.

\subsection{Vascular image synthesis}
Generative models, encompassing GAN \citep{goodfellow2014generative}, VAE \citep{kingma2013auto}, and diffusion models \citep{ho2020denoising}, have made rapid advancements in the field of synthesizing highly realistic images. Particularly, diffusion models have garnered significant attention due to their exceptional performance in image generation quality \citep{rombach2022high}.

 \citet{kim2022diffusion} proposed a Diffusion Adversarial Representation Learning (DARL) method for self-supervised vascular segmentation aimed at diagnosing vascular diseases. DARL comprises a diffusion module specifically designed to learn the distribution of background images and a generative module for vascular segmentation. \citet{kreitner2024synthetic} introduced a retinal vascular network simulation technique integrated with space colonization algorithms and three contrast adaptation pipelines. This method aims to enhance the accuracy of vascular segmentation by generating more realistic Optical Coherence Tomography Angiography (OCTA) synthetic images. \citet{lin2023yolocurvseg} proposed the YoloCurvSeg framework, which includes background generation, a vascular generator based on space colonization algorithms, and a multi-layer patch-wise contrastive learning synthesizer. This framework requires only a single noisy skeleton annotation to generate high-quality synthetic vascular datasets, significantly improving the accuracy of vascular segmentation. \citet{galati2023a2v} developed A2V for angiography-to-venography translation using StyleGAN2 architecture in brain vessel segmentation. \citet{feldman2025vesselgpt} introduced VesselGPT, an autoregressive approach combining VQ-VAE and GPT-2 to generate vessel sequences with B-spline-based cross-sectional representation.

Physics-based vascular synthesis methods \citep{jessen2025optimizing, jessen2023branching, ebrahem2024connecting}   typically employ nonlinear optimization frameworks, incorporating biophysical principles such as Murray's law and porous elasticity theory to generate vascular tree structures, requiring configuration of specific parameters including organ geometry, blood viscosity, and pressure drop constraints.  For example, \citet{jessen2025optimizing}  established an optimization framework based on Murray's law to generate multiple non-intersecting vascular trees in non-convex organs.

\subsection{Few-shot semantic segmentation}
Unlike ordinary natural images, vascular images are costly to acquire, and require expert knowledge for annotation, and their slender and complex structures often lead to high annotation costs, resulting in limited data availability \citep{cheng2024few}. To tackle annotation scarcity, one approach focuses on improving annotation efficiency through collaborative frameworks, while another leverages few-shot learning to train models with minimal labeled data.

For annotation efficiency, \citet{falcetta2025vesselverse} introduced VesselVerse, a large-scale brain vessel annotation dataset with 950 images from three public datasets across multiple neurovascular imaging modalities, featuring multi-expert annotations and version control with STAPLE-based consensus generation to facilitate collaborative refinement and quality assurance.

Few-shot Semantic Segmentation (FSS) methods offer an alternative solution by learning from limited annotations. FSS issues are commonly tackled by i) augmenting training data, ii) limiting the hypothesis space, or iii) adjusting the search strategy within the hypothesis space guided by prior knowledge extracted by a meta-learner \citep{song2023comprehensive}. Additionally, existing domain adaptation methods \citep{gu2022contrastive, galati2024federated, lin2024unsupervised, peng2022unsupervised, HU2024103164} typically handle cross-modality segmentation by developing domain-specific components or alignment strategies that minimize distributional differences between imaging modalities, utilizing techniques such as style transfer, contrastive learning, or multi-center collaborative training. For example,  \citet{HU2024103164} propose a Vector Field Transformer (VFT) that converts vessel images to vector fields using Hessian matrix eigenvectors and employs vessel enhancement networks to generate synthetic domains for cross-modality generalization between fundus photography and OCT angiography.

SSL leverages unlabeled data to enrich the training dataset, thereby assisting the learning of segmentation models.  \citet{chen2019multi, chen2021semi}  introduced SSL methods named MASSL and CPS, which concurrently optimize both supervised segmentation and unsupervised reconstruction objectives. However, the limited availability of unlabeled data in semi-supervised learning, caused by the difficulty in obtaining medical data, collectively leads to a bottleneck in performance improvement for the vascular segmentation task.

Meta-learning represents another approach to tackling FSS, featuring optimization-based, model-based, and metric-based methodologies \citep{raghu2019rapid, wang2021meta, snell2017prototypical}. Among these, metric learning has emerged as a leading paradigm for addressing few-shot segmentation challenges, as demonstrated by methods such as CANet, PANet, and PFENet \citep{zhang2019canet, wang2019panet, tian2020prior}. These approaches generally start by mapping support and query images into embedding features, followed by performing segmentation on the unseen query image through feature matching in the embedding space.

\section{Methodology}
\label{sec:methodology}
 The pseudo-vascular structure generation using the Spatial Colonization Algorithm (SCA) will be detailed in Sec.\ref{sec:Section E}. The formulation of a generative data-engine foundation model will be introduced in Sec. \ref{sec:Section A}. The compositional learning for flexible and diverse vascular synthesis will be presented in Sec. \ref{sec:Section B}. The few-shot generative adaptation method for transferable vascular image segmentation will be discussed in Sec. \ref{sec:Section C}. The collected UniVG-58K dataset will be introduced in Sec. \ref{sec:Section D}.

\subsection{Implementation details of the spatial colonization algorithm}
\label{sec:Section E}
\begin{figure}[h!]
    \centering
    \includegraphics[width=\linewidth]{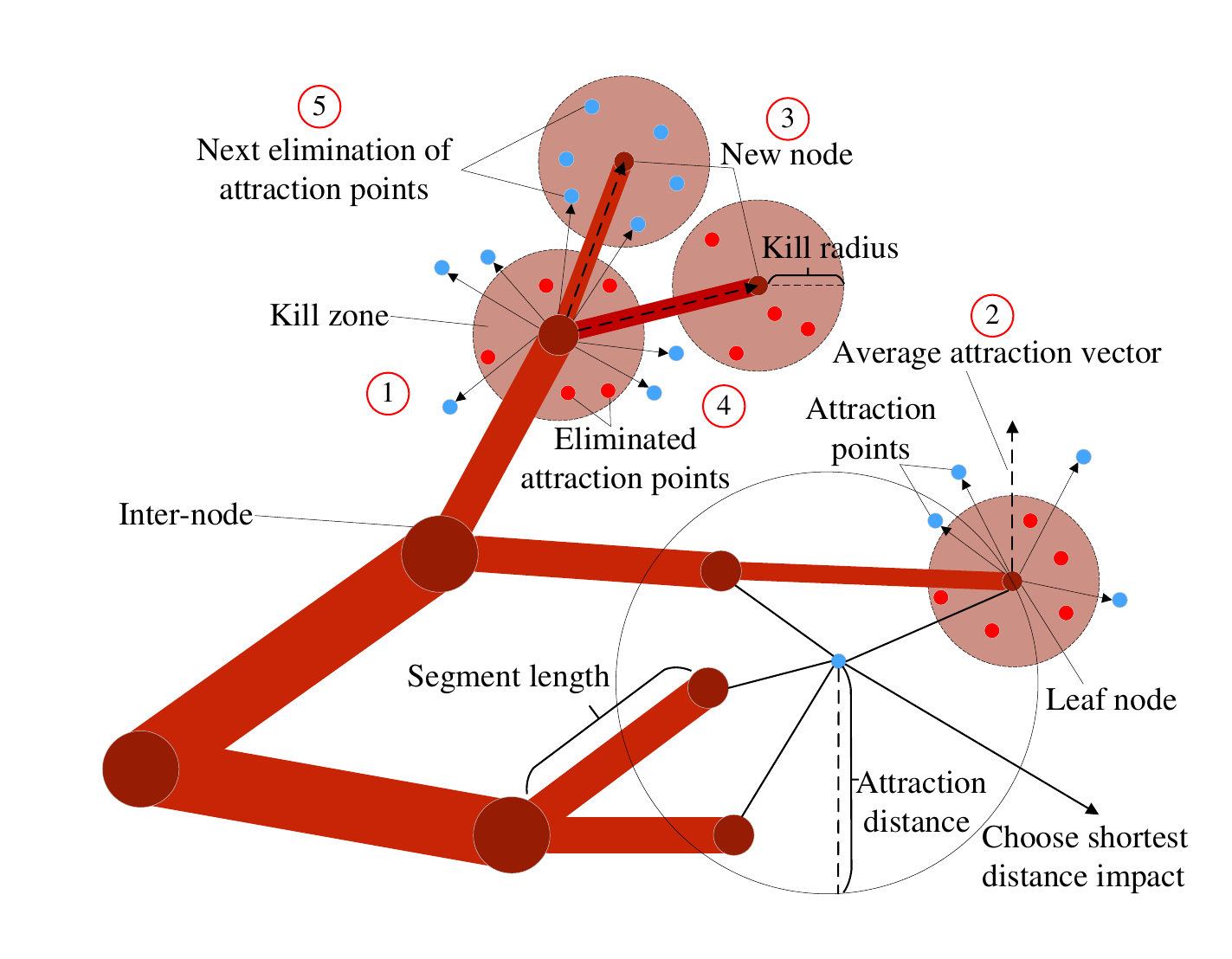}
    \caption{The process of node growth in the space colonization algorithm. Figure  details the direction of node growth, determined by the mean attraction vector. It also shows the next eliminated attraction points, the lethal radius, and the kill zone. }
    \label{fig:space colonization}
\end{figure}

For the vascular synthesis algorithm, we adopt the SCA proposed by \citep{runions2005modeling,runions2007modeling} is a procedural modeling algorithm used to simulate the growth of branching networks or tree-like structures, widely applied in the generation of vascular systems, leaf venations, root systems, and more. In UniVG, the algorithm flow is shown in Figure.\ref{fig:space colonization}, which simulates the iterative growth of curvilinear structures through attractors and nodes.

Nodes represent discrete points along vessel structures that progressively extend to form complete vascular networks, while attractors are spatial points that guide vessel growth direction by exerting influence on nearby nodes within a defined attraction distance (Da). We have employed a real-conditioned grid-based jittering approach for attractor placement, where the spatial distribution has been conditioned on real vascular masks from corresponding modalities with attractors placed using regular grids perturbed by random offsets to balance anatomical plausibility with structural diversity, and root nodes have been initialized at anatomically appropriate locations: optic disc centers for fundus imaging, multiple root nodes distributed across the foveal avascular zone boundary for OCT and OCTA, coronary ostia for coronary artery DSA, and major vessel origins for brain DSA.

The core steps include:
\begin{enumerate}
\item[1)]  \textbf{Placing attractors and associating nodes:} Attractors are placed randomly or according to a predefined pattern within the growth domain, where $\mathbf{a}_j$ represents the position of attractor $A_j$ and $\mathbf{p}_i$ represents the position of node $N_i$. These form the attractor set $\mathcal{A}$ and node set $\mathcal{N}$, respectively.

\item[2)] \textbf{Calculating the average direction of nodes:}
For each node, we determine the growth direction by averaging vectors pointing toward all attractors within the attraction distance $D_a$. The direction vector from node $N_i$ to each attractor $A_j$ within range is given by:
\begin{equation}
\mathbf{v}_{ij} = \mathbf{a}_j - \mathbf{p}_i, \quad |\mathbf{v}_{ij}| \leq D_a
\end{equation}
where $\mathbf{v}_{ij}$ is the direction vector from node $N_i$ to attractor $A_j$.
The average direction vector is then computed as:
\begin{equation}
\overline{\mathbf{v}}_i = \sum_{a_j \in \mathcal{A}} \mathbf{v}_{ij}
\end{equation}
where $\overline{\mathbf{v}}_i$ represents the average direction vector for node $N_i$.

\item[3)] \textbf{Calculating the position of new nodes:}
New nodes are generated by extending existing nodes along their computed growth directions. The average direction vector $\overline{\mathbf{v}}_i$ is first normalized to obtain a unit vector:
\begin{equation}
\hat{\mathbf{v}}_i = \frac{\overline{\mathbf{v}}_i}{\|\overline{\mathbf{v}}_i\|}
\end{equation}
where $\hat{\mathbf{v}}_i$ is the unit direction vector for node $p_i$.

The position of the new node $p_{i+1}$ is then calculated by advancing from the current node position by the segment length $L_s$:
\begin{equation}
\mathbf{p}_{i+1} = \mathbf{p}_i + L_s \cdot \hat{\mathbf{v}}_i
\end{equation}
where $\mathbf{p}_{i+1}$ is the position of the newly generated node.

\item[4)] \textbf{Checking the position of nodes:}
After placing new nodes at their calculated positions, we evaluate whether any attractors fall within the kill zone of the newly generated nodes. For each new node $p_{i+1}$ and each attractor $A_j$, we compute the distance:
\begin{equation}
d_{(i+1)j} = \|\mathbf{p}_{i+1} - \mathbf{a}_j\|
\end{equation}
where $d_{(i+1)j}$ is the distance between the new node $N_{i+1}$ and attractor $A_j$.

\item[5)] \textbf{Pruning attractors:}
Attractors within the kill distance of any newly added node are removed from the system. The updated attractor set is:
\begin{equation}
\mathcal{A} = \{a_j \in \mathcal{A} | d_{(i+1)j} > D_k\}
\end{equation}
where $\mathcal{A}$ is the updated attractor set, $d_{(i+1)j}$ is the distance between the new node $N_{i+1}$ and attractor $A_j$, and $D_k$ is the kill distance threshold.

\item[6)] \textbf{Iteration and termination:}
The algorithm iteratively repeats steps 1)--5) until one of the termination conditions is satisfied. The algorithm terminates when:
\begin{equation}
|\mathcal{N}| \geq N_{\max} \quad \text{or} \quad t \geq T_{\max}
\end{equation}
where $N_{\max}$ is the maximum allowed number of nodes, $T_{\max}$ is the maximum number of iterations, and $t$ is the current iteration count.
\end{enumerate}

As shown in the spatial colonization algorithm in Fig.\ref{fig:space colonization}, the process involves several key steps to generate diverse vascular structures. By observing the starting point, boundary, and curvature of vascular structures in images, parameters such as the root node coordinates \( C_r \) (e.g., the starting point of retinal vascular located in the optic disc region) as well as boundaries and obstacles can be set. Commonly used parameters such as attraction distance \( D_a \), kill distance \( D_k \), and segment length \( L_s \) are set to  20, 10, and 15, respectively, but can be adjusted as needed. The number of attractors is controlled through a grid placement strategy, and jitter is introduced to increase randomness. Additionally, by recursively calculating the radius from the branch tips to the tree base, the thickness variation of branches is simulated. Finally, by setting random attractors and root nodes, a library of curves with varied shapes but the same type is generated for pre-training the mask generation model.
\subsubsection{Spatial colonization algorithm initialization}
\textbf{S}patial \textbf{c}olonization \textbf{a}lgorithm initialization incorporates \textbf{r}eal vessel topology, denoted as R-SCA, to enhance structural authenticity. This algorithm comprises two sequential steps:
\begin{enumerate}
\item[1)] \textbf{Real image initialization:} The algorithm employs a real-conditioned hybrid placement approach combining dense vessel coverage with randomized grid-based distribution. The spatial distribution is conditioned on real vascular masks from corresponding modalities, where attractors are first densely placed at true vessel locations to ensure anatomical fidelity. In the remaining non-vessel regions, a regular grid is established and attractors are randomly placed within grid cells to introduce spatial variation and placement diversity. This hybrid strategy ensures both adherence to anatomical vascular territories and sufficient structural variation for diverse pseudo-vascular generation.\item[2)] \textbf{Spatial colonization-based structure generation:} The spatial colonization algorithm is employed to generate vessel masks by iteratively growing branches from initialized root nodes toward attractors.
\end{enumerate}

\subsection{Formulation of vascular data-engine foundation model}
\label{sec:Section A}

A foundational generative data-engine model for universal few-shot vascular image segmentation generates a large quantity of high-quality and diverse vascular images from limited annotated data, thereby driving vascular image segmentation algorithms. Specifically, the framework comprises three components: the vascular mask data-engine (detailed in Sec. \ref{sec:Vascular mask data-engine}), the vascular image data-engine (introduced in Sec. \ref{sec:Vascular image data-engine}) and the vascular data-engine  (explained in Sec. \ref{sec:Vascular image data-engine}).

\subsubsection{Vascular mask data-engine}
\label{sec:Vascular mask data-engine}

R-SCA (referred to as $S$) generates a large quantity of synthetic vascular labels (denoted by the set $\mathcal{M_F}$, where each element is $m_i$). The formulation is as follows,

\begin{equation}
\mathcal{M_F} = \{m_i | m_i = S(\mathbf{I}, \mathbf{P}), \forall i \in \mathbb{N}\},
\end{equation}
where $\mathbb{N}$ represents the set of natural numbers, indicating that the index $i$ runs over all instances of synthetic vascular labels. $S(\cdot)$ processes the input image $\mathbf{I}$ to generate each synthetic vascular label $m_i$, $\mathbf{P}$ represents the set of parameters used for generating synthetic vascular labels.

The vascular mask data-engine is trained using a large number of pseudo-vascular labels $\mathcal{M_F}$.  This learning process is formalized as,

\begin{equation}
     \theta_1=G_M(\mathcal{M_F}| \theta),
    \label{eq: pretrained}
\end{equation}
where $\theta$ represents the parameters of a mask data-engine pretrained on natural images, and $\theta_1$ denotes the parameters of the vascular mask data-engine after pretraining on a large-scale dataset of vascular structures.

\subsubsection{Vascular image data-engine} 
\label{sec:Vascular image data-engine}
To generate synthetic vascular images that align with the corresponding vascular masks for downstream segmentation tasks, we use the vascular image data-engine ($G_I$), which  is trained on large unannotated vascular image dataset \( \mathcal{I} = \{I_i |  i \in \mathbb{N}\} \) to learn the background distribution within the vascular images.
\begin{equation}
     \theta_2=G_I(\mathcal{I} | \theta),
    \label{eq: pretrained}
\end{equation}
where $\theta$ represents the parameters of the vascular image data-engine, $\theta_2$ represents the parameters of the model after pre-training.

\subsubsection{Vascular data-engine} 
\label{sec.Vascular data engine}
Building upon the vascular mask data-engine and the vascular image data-engine, we obtain the  specialized vascular mask data-engine ($G_M$) and the  specialized vascular image data-engine (CondG)  after adaptation to specialized vascular domains (Sec. \ref{sec.efficient adaptation}). Combining these two engines results in the final data-engine. Using the synthetic vascular masks $S_{mask}$ generated by the  specialized vascular mask data-engine as described by:
\begin{equation}
S_{mask} = \{S_{mask}^i | S_{mask}^i = G_M(\theta_3 , \epsilon_i), i \in \mathbb{Z} \},
\label{eq:mask generator}
\end{equation}
where $\epsilon_i$ is the randomly sampled noise and $\theta_3$ is defined in Sec. \ref{sec.efficient adaptation}.

The  CondG is utilized to generate vascular images $S_{img}$
that align with the vascular masks and feature diverse backgrounds.
The formal description is given by:
\begin{equation}
S_{img} = \{S_{img}^i | S_{img}^i = CondG(S_{mask}^i | \theta_4,  \epsilon_i), i \in \mathbb{Z}\},
\label{eq:vascular data engine}
\end{equation}
where $\epsilon_i$ is the randomly sampled noise and $\theta_4$ is defined in Sec. \ref{sec.efficient adaptation}.

\begin{figure*}[!t]
    \centering
    \includegraphics[width=\textwidth,height=0.6\textheight,keepaspectratio]{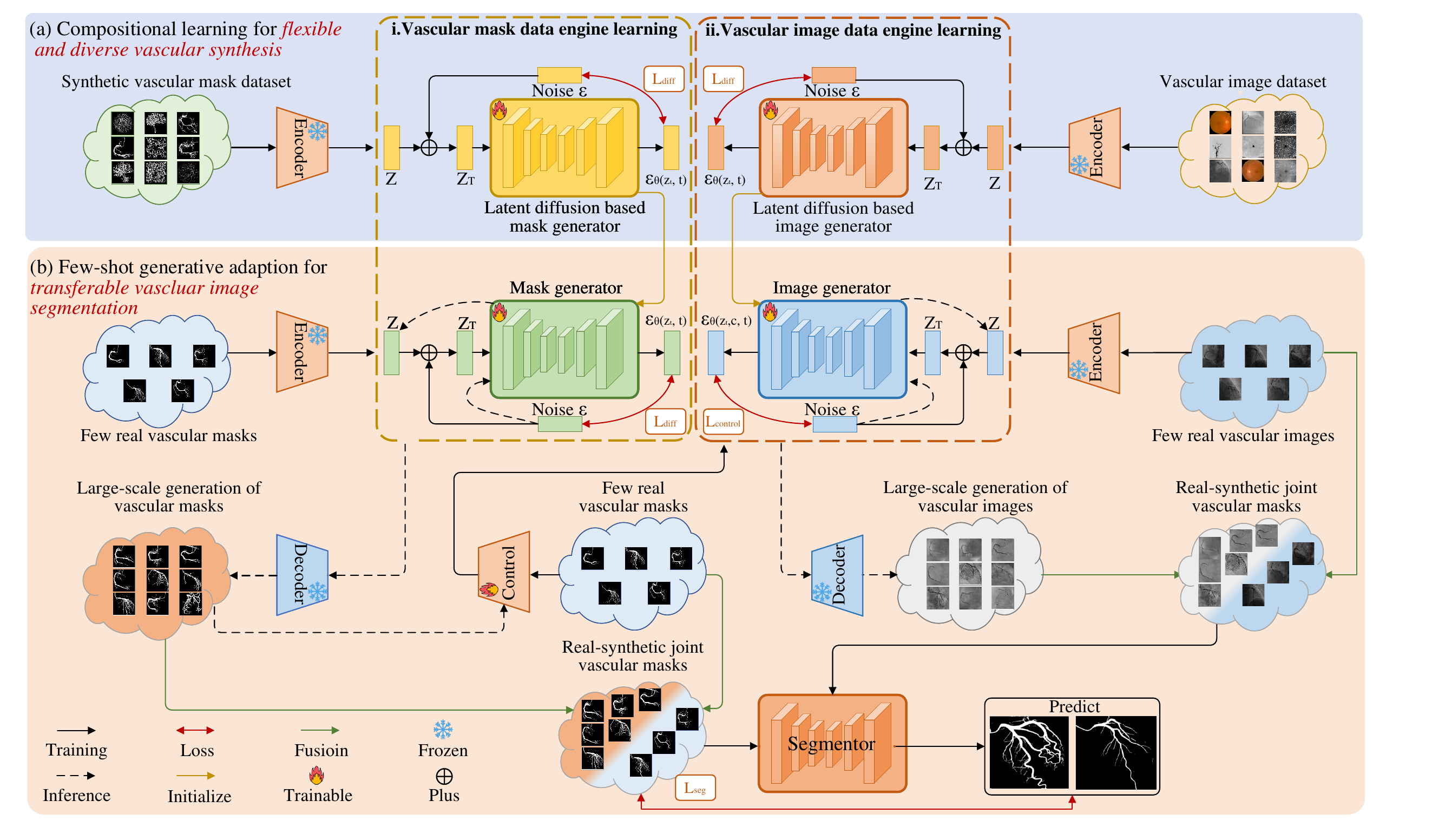}
    \vspace{-1cm} \caption{UniVG framework architecture overview. The framework consists of two stages: (a) Compositional learning for flexible and diverse vascular synthesis, including i) vascular mask data-engine learning using spatial colonization algorithms and latent diffusion models to learn morphological knowledge, and ii) vascular image data-engine learning from the UniVG-58K dataset for background distribution modeling. (b) Few-shot generative adaptation using 5-shot real samples to fine-tune both generators, bridging the synthetic-real domain gap for downstream segmentation tasks.}
    \label{fig:framework}
\end{figure*}

\subsection{Compositional learning for flexible and diverse vascular synthesis}
\label{sec:Section B}
\subsubsection{Learning the vascular mask data-engine from large-scale synthetic vascular data}

The vascular mask data-engine learns diverse vascular morphological structures from a large number of pseudo-labels, enabling the model to synthesize a variety of vascular structures. Our approach consists of two components (Fig. \ref{fig:framework} i): 1) A regularization algorithm for vascular structure synthesis: We propose a novel vascular structure synthesis algorithm that uses spatial colonization algorithms (SCA) \citep{runions2005modeling,runions2007modeling} to randomly place a large number of attractors conditioned on real vessels. This approach guides the growth of vascular structures following the SCA growth pattern, enabling the synthesis of a wide range of vascular structures across different anatomical sites. The final output is a synthetic vascular mask dataset ($\mathcal{M_F}$).
2) Learning the vascular mask data-engine: To enable efficient learning of the vascular mask data-engine, we fine-tune the pre-trained Stable Diffusion model using Low-Rank Adaptation (LoRA) \citep{hu2021lora}. This technique allows the model to transfer knowledge from natural images to vascular structures, enhancing its ability to synthesize high-quality vascular masks. Firstly, the synthetic vascular mask $m_i$ is transformed into a latent variable $Z$ through an encoder $E(\cdot)$. Then, at each time step $t$, noise $\epsilon$ is added to the latent variable $Z$ to generate the noisy latent variable $Z_t$:

\begin{equation}
    Z_t = \sqrt{1 - \beta_t} Z + \sqrt{\beta_t} \epsilon, \quad
    \label{equ.8}
\end{equation}
where $Z = E(m_i)$ is the latent variable obtained by encoding the synthetic vascular
mask $m_i$. $\beta_t$ is the variance scheduling parameter used to control the amount of noise added at each step. $\epsilon \sim \mathcal{N} (0, I)$ is the noise sampled from a standard normal distribution.

Next, the vascular mask data-engine $\epsilon_\theta(\cdot)$ predicts the noise $\epsilon _\theta(Z_t, t)$ in the noisy latent variable $Z_t$, and the difference between the predicted noise and the actual noise is calculated as the loss function:
\begin{equation}
    L_{\text{diff}} = \mathbb{E}_{z, t, \epsilon} \left[ \| \epsilon - \epsilon _\theta(Z_t, t) \|_2^2 \right], \quad
    \label{equ.9}
\end{equation}
where  $\epsilon \sim \mathcal{N}(0, I)$ is the noise sampled from a standard normal distribution. The time step $t$ is sampled from a uniform distribution $U(1, T)$.

\subsubsection{Learning the vascular image data-engine from large-scale vascular images}

The vascular image data-engine (Fig. \ref{fig:framework} ii) achieves an efficient representation of vascular images by learning extensively from a wide range of 2D vascular images $\mathcal{I}$ (Sec. \ref{sec:Section D}) spanning multiple modalities and categories. This extensive learning enables the engine to perform compositional generation of both vascular structures and vascular backgrounds. Firstly, the vascular image $I_i$ is transformed into a latent variable $Z$ through an encoder $E(\cdot)$, obtaining a noisy latent variable $ Z_t = E(I_i) \quad$(Equ. \ref{equ.8} ). Then, the noise $\epsilon_\theta(Z_t, t)$ is predicted by the vascular image data-engine and the loss is calculated using $L_{\text{diff}} = \mathbb{E}_{z, t, \epsilon} \left[ \| \epsilon - \epsilon_\theta(Z_t, t) \|_2^2 \right] \quad$ (Equ. \ref{equ.9}).

\subsubsection{Hierarchical compositional mechanism}

Our compositional learning has operated through a hierarchical decomposition and recombination process that has been embedded throughout the UniVG architecture. The mechanism has functioned at three distinct levels:

\textbf{a) Structural decomposition}: Structural decomposition has been achieved via the iterative denoising process, where the diffusion model effectively resolves vascular masks into fundamental elements (e.g., vessel segments, bifurcation points). These decomposed elements have been encoded as hierarchical latent features within the diffusion model's learned representations, where lower-level features capture local geometric primitives such as vessel curvature and thickness, while higher-level features represent global topological patterns including branching connectivity and overall network architecture. This hierarchical feature encoding has enabled the model to capture essential morphological characteristics across different vessel types and anatomical contexts.

\textbf{b) Feature recombination}: The diffusion-based vascular mask data-engine has learned to recombine these decomposed structural elements in novel configurations, utilizing structural features decomposed from fine-tuned real masks as anchor points. The model has balanced synthetic diversity with anatomical realism by composing new vascular structures from learned primitives. 

\textbf{c) Cross-modal composition:} The vascular image data-engine has composed the corresponding modality-specific vessel appearances by combining structural masks with diverse background contexts. This compositional process has enabled the generation of morphologically diverse vascular images that have maintained anatomical plausibility across different imaging modalities.

\textbf{Summary of advantages:}  \textbf{1) Flexibility:} 
The vascular mask data-engine possesses the capability to customize the complexity and layout of vascular structures by fine-tuning the vasculature in the corresponding domain. This engine supports the generation of multiple types of vascular structures and can flexibly integrate background information with a diverse array of vascular masks, thereby creating images that are stylistically unique and content-rich. \textbf{2) Diversity:} By acquiring multi-level knowledge from large-scale collections and generated vascular images, this approach enables a broader and more complex combination of vascular structures. Consequently, it facilitates the generation of morphologically diverse vascular images using only a minimal amount of labeled data.

\subsection{Few-shot generative adaptation for transferable vascular image segmentation}
\label{sec:Section C}
\subsubsection{Efficient adaptation for  specialized vascular images}
\label{sec.efficient adaptation}
Efficient adaptation for specific vascular images enables the specialization of pre-trained knowledge to specific vascular distributions, thereby generating diverse vascular structures from limited annotated data.
Including two parts of downstream adaptation:

\textit{1) Specialized vascular mask data-engine specific to downstream task:} We fine-tune the vascular mask data-engine using Low-Rank Hadamard Product (LOHA) \citep{hyeon2021fedpara} techniques with a small dataset of real vascular masks corresponding to specific anatomical locations. The weight matrix ($W'$) after LoRA adjustment becomes:
\begin{equation}
W' = W + \lambda_1 \Delta W,
\label{equation: Hyperparameters}
\end{equation} 
where $\lambda_1$ is used to control the influence strength of the adjustment term $\Delta W$ on the original weight matrix $W$. Using five real and domain-specific vascular labels $R_{mask} = \{ R_{mask}^i \mid i \in \mathbb{Z}, 1 \leq i \leq 5 \}$ as the training set, the vascular image $m_i$ is transformed into a latent variable $Z$ through an encoder $E(\cdot)$, obtaining a noisy latent variable $Z_t \quad$ (Equ. \ref{equ.8}). Then, the noise $\epsilon_\theta(Z_t, t)$ is predicted by the vascular image data-engine and the loss is calculated using $L_{\text{diff}}$ (Equ. \ref{equ.9}). After training, we obtain the specialized mask data-engine ($G_M$):

\begin{equation}
\theta_3 = G_M(R_{mask} \mid  \theta_1),
\label{eq:mask generator}
\end{equation}
where  $\theta_3$ denotes the parameters of the generative model after fine-tuning.

\textit{2) Specialized vascular image data-engine specific to downstream task: } Based on the vascular image data-engine, we use five labeled vascular images $R_{img} = \{ R_{img}^i \mid i \in \mathbb{Z}, 1 \leq i \leq 5 \}$ and their corresponding masks $R_{mask}$ as the training set. The vascular image $R_{img}^i$ is transformed into a latent variable $Z$ through an encoder $E(\cdot)$. The vascular mask $R_{mask}^i$ is downsampled to the same size as $Z$ using a conditional network to obtain $c_f$. A noisy latent variable $Z_t$ is obtained through Equ. \ref{equ.8}. The noise is then predicted using a noise prediction model $\epsilon_\theta(Z_t, t, c_f)$, and the loss is calculated using $L_{\text{control}}$:
\begin{equation}
    \mathcal{L}_{control} = \mathbb{E}_{Z_t, t, c_f, \epsilon \sim \mathcal{N}(0,1)} \left[ \| \epsilon - \epsilon_\theta(Z_t, t, c_f)) \|_2^2 \right],
\end{equation}
After the training process is completed, we obtain the  specialized vascular image data-engine (denoted as CondG). This process can be described as follows:
\begin{equation}
    \theta_4 = CondG(R_{img}, R_{mask} | \theta_2),
    \label{eq:vascular data engine}
\end{equation}
where $\theta_4$ represents the parameters of the model after fine-tuning.

\subsubsection{Downstream vascular image segmentation}
The vascular data-engine (Sec. \ref{sec.Vascular data engine}) was utilized to generate a large-scale dataset comprising vascular masks $S_{mask}$ alongside their corresponding vascular images $S_{img} $.  In addition to the generated dataset, a dataset containing 5 real vascular images and masks was also prepared. These two datasets are jointly used for training downstream segmentation tasks. 
Specifically, during the training phase, to ensure authenticity during the learning process and avoid interference from deviations introduced by synthetic data in segmentation, we randomly select a synthetic vessel image ($S_{img}^i$) and its corresponding mask ($S_{mask}^i$) from the synthetic training image set, and a real vessel image ($R_{img}^i$) and its corresponding mask ($R_{mask}^i$) from the real fine-tuning vessel dataset. These synthetic and real vessel images are then placed into a batch as input to the downstream segmentation network, with the synthetic and real masks serving as training targets. Here, $pred(\cdot)$ denotes the segmentation mask predicted by the model based on the input image (whether synthetic or real). The loss function uses binary cross-entropy loss (BCE), defined as:
\begin{equation}
\begin{split}
L_{seg} = & BCE(S_{mask}^i, pred(S_{img}^i)) \\
       & + BCE(R_{mask}^i, pred(R_{img}^i)).
\end{split}
\end{equation}

By leveraging the aforementioned training strategy, the network can learn rich and diverse features from large-scale synthetic data while using a small amount of real data to ensure generalization to real-world scenarios. This approach enables the model to achieve highly accurate vessel image segmentation even when only a limited number of annotated real images are available.

\begin{figure}
    \centering
    \includegraphics[width=\linewidth]{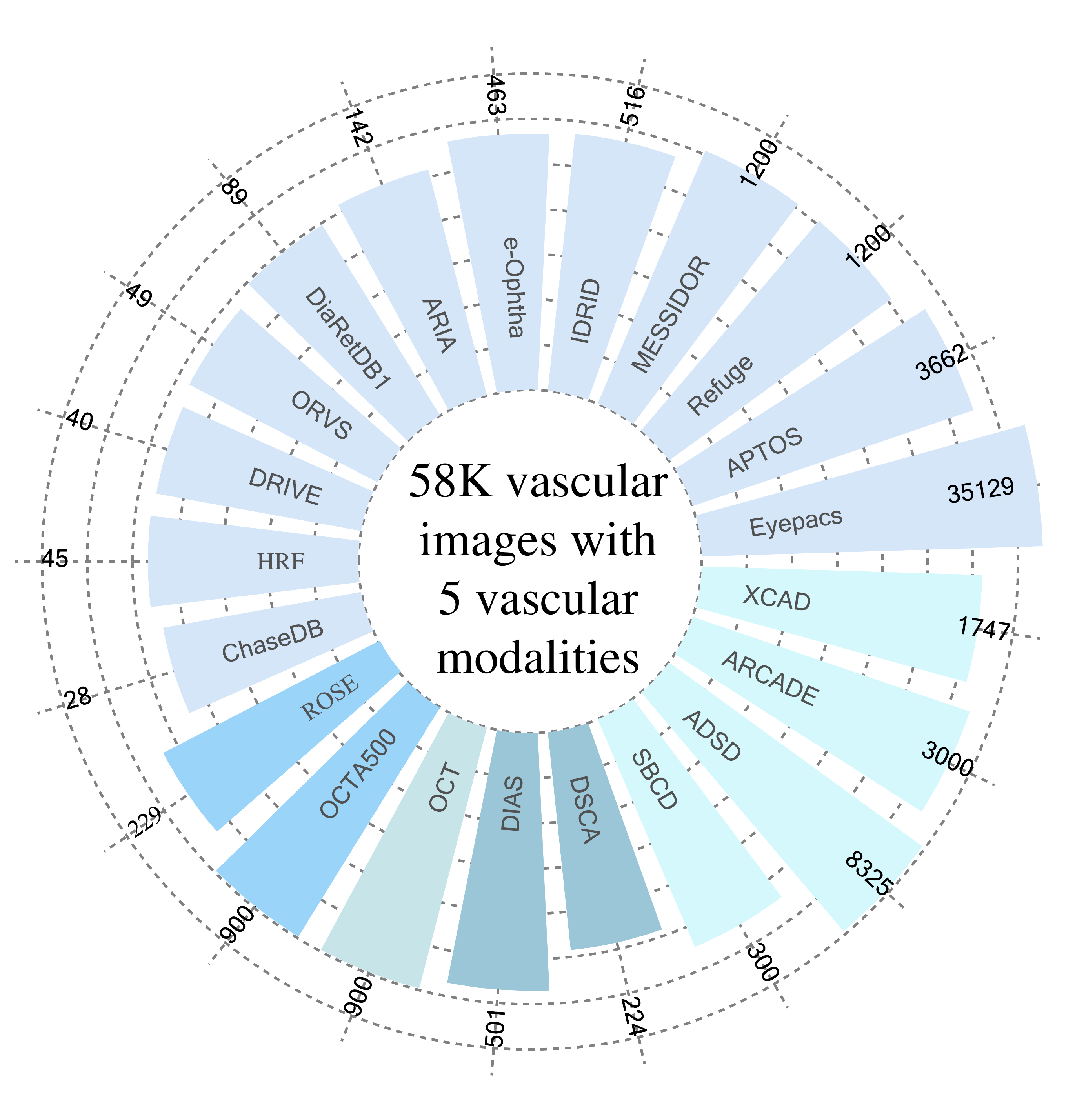}
    \caption{We curated a large-scale 2D vascular dataset, UniVG-58K, for pre-training and utilized it as the downstream dataset.}
    \label{fig:large_scale_iamges}
\end{figure}

\textbf{Summary of advantages:}  \textbf{1) Transferability:} Efficient adaptation for specialized vascular images exhibits remarkable versatility, requiring only 5 images and corresponding masks to be applied across various segmentation tasks. This advantage is attributed to its few-shot generation adaptation capability, enabling rapid deployment in different vascular domains. \textbf{2) Accuracy:} By generating diverse and realistic images, the accuracy and reliability of segmentation outcomes are significantly enhanced, providing high-quality data support for downstream tasks.

\subsection{UniVG-58K dataset}
\label{sec:Section D}

We have curated a large dataset named UniVG-58K (Fig. \ref{fig:large_scale_iamges}), which encompasses 58,689 vascular images across five distinct imaging modalities: Optical Coherence Tomography (OCT), OCTA, Fundus imaging, Brain Digital Subtraction Angiography (DSA), and Coronary Artery DSA. This dataset integrates content from twenty publicly available datasets and one private dataset, ensuring a diverse representation of vascular structures. Specifically, the fundus imaging dataset \citep{kauppi2007diaretdb1, decenciere2013teleophta, porwal2018indian, decenciere2014feedback, orlando2020refuge, karthik2019aptos, gulshan2016development} includes 42,259 images from seven authoritative sources, while the coronary artery DSA dataset \citep{Danilov2021AngiographicDF, maxim2023arcade, ma2021self} comprises 12,946 images from three different collections. Additionally, the cerebral DSA dataset \citep{liu2024dias}utilizes the DIAS dataset with 441 images. The OCT and OCTA datasets \citep{li2024octa, ning2024accurate} contribute an additional 600 and 829 high-resolution images, respectively. The dataset is publicly available at https://huggingface.co/datasets/xinaloha/UniVG. Detailed descriptions of these datasets are provided in the  \ref{appendix:More details of UniVG datasets} for further reference.

The UniVG-58K dataset is meticulously designed to enhance the performance of foundational models in few-shot medical image segmentation tasks. By covering multiple imaging modalities, it provides rich training materials that help improve the model's understanding and recognition capabilities for various vascular structures. The diversity and scale of the dataset enable foundational models to learn more representative features, thereby exhibiting stronger adaptability when handling complex and diverse vascular images.

\section{EXPERIMENTS CONFIGURATIONS}

\subsection{Experiments tasks settings}
The UniVG-58K dataset is split into one part for foundation model pretraining and another for few-shot vascular segmentation in downstream tasks.

1) For foundation model pretraining:  The pre-training dataset contains a total of 57,075 images covering five modalities: Fundus (42,259 images), Coronary artery DSA (12,946 images), Brain DSA (441 images), OCTA (829 images), and OCT (600 images).

2) For few-shot vascular segmentation in downstream tasks: a) Few-shot fundus vascular segmentation (FVS):
This task focuses on vascular segmentation within fundus images, incorporating datasets such as DRIVE (40 images) \citep{asad2014ant}, CHASEDB1 (28 images) \citep{henry2020mixmodule}, HRF (45 images) \citep{li1903connection}, ORVS (49 images) \citep{sarhan2021transfer}, and ARIA (142 images) \citep{bankhead2012fast}. 
b) Few-shot coronary artery DSA segmentation (FCS): This task targets coronary artery segmentation using digital subtraction angiography (DSA) images, including datasets like XACD (126 images) \citep{ma2021self} and SBCD (300 images).  c) Few-shot brain DSA segmentation (FBS): The brain DSA image segmentation task includes datasets DIAS (60 images) \citep{liu2024dias} and DSCA (224 images) \citep{xie2024dsnet, xie2024dsca}.  
d) Few-shot OCTA vascular segmentation 
 (FOS): Optical coherence tomography angiography (OCTA) vascular segmentation utilizes the OCTA500 dataset (300 images) \citep{li2024octa}.  e) Few-shot OCT vascular segmentation (FTS):
The final task involves optical coherence tomography (OCT) vascular segmentation using another OCT dataset (300 images) \citep{li2024octa}. Similar to OCTA, this dataset is used for studying detailed structures of the eye, particularly the retina. 

The aforementioned five tasks follow this protocol: The numbers in parentheses indicate the number of labeled images in each dataset. For training purposes, five randomly selected labeled images are used as the training set, while the remaining images serve as the test set. Specifically, the training and testing set splits for each dataset are presented in Table \ref{tab:pre-vascular}.

\subsection{Comparison settings} 
To demonstrate the superiority of our proposed UniVG framework, we compare it with 21 widely-used frameworks in the few-shot scenario. 1) Direct Learning Segmentation: The UNet \citep{ronneberger2015u} and nnUNet \citep{isensee2021nnu} are utilized to evaluate the performance of supervised medical image segmentation models in the 5-shot scenario. Both models are trained using the same 5-shot images and corresponding masks to ensure fair comparison under data-limited conditions. The UNet serves as a basic baseline to assess fundamental supervised learning performance, while nnUNet provides a more advanced baseline with automated architecture optimization and preprocessing strategies. 
Both methods employ five-fold cross-validation on the complete datasets to assess the upper limit of segmentation performance under fully supervised scenarios. Additionally, we compared two specialized vessel segmentation methods on DRIVE and CHASEDB datasets, including Retina-Unet \citep{jaeger2020retin} and FSG-Net \citep{bacchin2023fsg}, trained using the same 5-shot images and corresponding masks, following their original preprocessing and postprocessing protocols. 2) Weak Supervision Learning: We compare MASSL \citep{chen2019multi}, CPS \citep{chen2021semi}, CCT \citep{ouali2020semi}, AGMM \citep{wu2023sparsely}, DBDM \citep{luo2022scribble}, PVA \citep{zhang2023partial} and Vessel-captcha\citep{dang2022vessel}  to illustrate the limitations posed by the lack of domain knowledge in the few-shot scenario and the noise-induced weak supervision challenges during training. 3) Pre-trained Foundation Model: A series of pre-trained foundation models (MAE \citep{he2022masked}, SimCLR \citep{liu2022deep}, MOCO \citep{he2020momentum}, iBOT \citep{zhou2021ibot}, SimSiam \citep{chen2021exploring}, SAM \citep{kirillov2023segment}, MedSAM \citep{ma2024segment}) are compared to highlight the limitations encountered when transferring pre-trained knowledge to medical images due to domain gaps. 4) Pre-trained Foundation Models on Vascular Images (PFM-V): Models such as iBOT \citep{zhou2021ibot} and SimSiam \citep{chen2021exploring} are compared to showcase the challenges associated with transferring contrastive learning-based pre-training paradigms to vascular images.To ensure fair comparison, we have adapted the original pre-trained models from both methods to our vascular imaging domain while preserving their respective backbone architectures. For iBOT, we utilized their Vision Transformer-Small/16 (ViT-S/16) architecture following their self-distillation approach with masked image modeling framework. For SimSiam, we employed their ResNet-50 backbone with the characteristic stop-gradient operation and Siamese network design. Both methods were pre-trained on our large-scale vascular dataset using publicly available ImageNet pre-trained checkpoints, adhering to their original training protocols and hyperparameter configurations.

5) Data Engine Paradigm (DE): We compare YoloCurvSeg \citep{lin2023yolocurvseg} to evaluate the performance of skeleton-based synthetic data generation approaches in cross-modal vascular segmentation scenarios. To ensure fair comparison, we provided YoloCurvSeg with 20 images containing skeleton noise annotations for each vascular modality and followed their original training protocols with all network hyperparameters maintained according to the paper's configuration. We also compared with the SOCT \citep{kreitner2024synthetic} method, generating 500 OCTA vascular images by fully following the SOCT \citep{kreitner2024synthetic} configurations, using the same UNet network and training parameters for testing on the OCTA500 dataset.

\subsection{Implementation details}
All training processes were executed on NVIDIA GeForce RTX V100 GPUs with 32GB memory. 
\textit{1) For the vascular mask data-engine learning:} Pseudo vascular structures totaling 10,000 (2,000 per modality across fundus imaging, OCTA, OCT, brain DSA, and coronary artery DSA) were generated using SCA and used for training on an initialized Stable Diffusion v1.5 model\footnote{\url{https://huggingface.co/stable-diffusion-v1-5/}}. The learning rate was set at $1\times10^{-4}$ with a total of  $6\times10^{5}$ iterations. 
\textit{2) Specialized vascular mask data-engine specific to downstream task: }Fine-tuning was performed on the vascular mask data-engine using the LOHA method, maintaining the same learning rate of
$1\times10^{-4}$. A cosine learning rate scheduler was employed, with convolutional dimension and convolutional alpha both set to 4. The batch size was set to 1, with fine-tuning iterations limited to 1,500. Both pre-training and fine-tuning stages utilized the Adam optimizer, with network dimensions and network alpha set to 32. The hyperparameter $\alpha_1$ for generating vascular masks was set to 0.7.
\textit{3) For the vascular image data-engine learning:} An initialized Stable Diffusion v1.5 was pretrained on the UniVG-58K dataset with a learning rate of $1\times10^{-4}$, using the Adam optimizer, and a batch size of 8 for  $1.4\times10^{5}$ iterations.
\textit{4) Specialized vascular image data-engine specific to downstream task:} The vascular image data-engine served as the foundation for ControlNet training, which was further trained using 5 annotated vascular images. The learning rate for this stage was reduced to $1\times10^{-5}$, with a batch size of 2 and $6\times10^{3}$ iterations.
\textit{5) For downstream vascular image segmentation:} For downstream vascular segmentation tasks, a standard UNet architecture was selected. The loss function was defined as cross-entropy loss, optimized with the Adam optimizer, starting with an initial learning rate of $3\times10^{-4}$. The training process was capped at a maximum of 300 epochs. Notably, all generated images were used directly without any filtering or quality selection process, and no post-processing steps were applied after segmentation.

\section{RESULTS AND ANALYSIS}

\begin{table*}
\centering
\caption{Quantitative evaluation demonstrates the superiority of our UniVG framework in few-shot medical segmentation tasks. Compared to 22 methods, our UniVG framework achieves the best average performance across 11 tasks and 5 vascular modalities. N/A entries marked with "\textemdash".}
 \label{tab: comparative experiment}
    \begin{subtable}{\textwidth}
    \centering
   
    \begin{adjustbox}{width=\textwidth,totalheight=0.95\textheight}

    \begin{tabular}{llcccccccc}
    \toprule
    \multirow{2}{*}{Model} & \multirow{2}{*}{Type}  & \multicolumn{2}{c}{FVS-DRIVE} & \multicolumn{2}{c}{FVS-CHASEDB1} &  \multicolumn{2}{c}{FVS-HRF} & \multicolumn{2}{c}{FVS-ORVS} 
    \\ 
    \cmidrule(lr){3-4} \cmidrule(lr){5-6} \cmidrule(lr){7-8} \cmidrule(lr){9-10}  
     & & DSC$\uparrow$ & clDice $\uparrow$ & DSC$\uparrow$ & clDice$\uparrow$ & DSC$\uparrow$ & clDice$\uparrow$ & DSC$\uparrow$ & clDice$\uparrow$  \\
    
    \midrule
     
\textcolor{gray}{UNet 
\citep{ronneberger2015u}} 
        & \textcolor{gray}{FIVE} 
        & \textcolor{gray}{$64.39_{\pm 6.00}$}  
        & \textcolor{gray}{$64.22_{\pm 6.05}$}  
        & \textcolor{gray}{$72.62_{\pm 3.04}$} 
        & \textcolor{gray}{$73.53_{\pm 3.75}$}
        & \textcolor{gray}{$63.18_{\pm 6.30}$}  
        & \textcolor{gray}{$59.03_{\pm 7.20}$}  
        & \textcolor{gray}{$47.80_{\pm 11.54}$} 
        & \textcolor{gray}{$42.32_{\pm 11.50}$}    \\

\textcolor{gray}{nnU-Net \citep{isensee2021nnu}}
        & \textcolor{gray}{FIVE} 
        & \textcolor{gray}{$79.12_{\pm 3.06}$}  
        & \textcolor{gray}{$81.02_{\pm 4.01}$}  
        & \textcolor{gray}{$84.26_{\pm 1.47}$} 
        & \textcolor{gray}{$87.05_{\pm 2.43}$}
        & \textcolor{gray}{$81.21_{\pm 4.33}$}  
        & \textcolor{gray}{$84.52_{\pm 4.47}$}  
        & \textcolor{gray}{$75.04_{\pm 5.32}$} 
        & \textcolor{gray}{$78.37_{\pm 5.21}$}    \\

\textcolor{gray}{Retina-Unet \citep{jaeger2020retin}}       & \textcolor{gray}{FIVE}  & \textcolor{gray}{$68.61_{\pm 3.49}$} & \textcolor{gray}{$69.49_{\pm 4.48}$}   & \textcolor{gray}{$70.92_{\pm 3.11}$}  &  \textcolor{gray}{$73.47_{\pm 3.21}$} & \textcolor{gray}{$74.77_{\pm 5.69}$}  & \textcolor{gray}{$75.55_{\pm 6.60}$}  & \textcolor{gray}{$65.15_{\pm 7.41}$} & \textcolor{gray}{$66.84_{\pm 7.43}$}  \\
\textcolor{gray}{FSG-Net \citep{bacchin2023fsg}} & \textcolor{gray}{FIVE}  & \textcolor{gray}{$77.90_{\pm 3.23}$} & \textcolor{gray}{$80.35_{\pm 4.10}$}   & \textcolor{gray}{$77.03_{\pm 1.98}$}  &  \textcolor{gray}{$78.51_{\pm 2.32}$} & \textcolor{gray}{$77.31_{\pm 4.64}$}  & \textcolor{gray}{$78.92_{\pm 4.83}$}  & \textcolor{gray}{$70.08_{\pm 6.03}$} & \textcolor{gray}{$71.11_{\pm 6.22}$}  \\
         \textcolor{gray}{UNet \citep{ronneberger2015u}} 
        & \textcolor{gray}{FULL}   
        & \textcolor{gray}{$79.24_{\pm 4.03}$}   
        & \textcolor{gray}{$81.40_{\pm 4.96}$} 
        & \textcolor{gray}{$80.55_{\pm 2.12}$}  
        & \textcolor{gray}{$84.52_{\pm 2.62}$}  
        & \textcolor{gray}{$74.14_{\pm 6.09}$}   
        & \textcolor{gray}{$73.05_{\pm 7.82}$}  
        & \textcolor{gray}{$74.66_{\pm 4.61}$} 
        & \textcolor{gray}{$77.80_{\pm 4.75}$} \\
       \textcolor{gray}{nnU-Net \citep{isensee2021nnu}}
        & \textcolor{gray}{FULL}   
        & \textcolor{gray}{$80.90_{\pm 1.55}$}   
        & \textcolor{gray}{$82.37_{\pm 3.76}$} 
        & \textcolor{gray}{$86.01_{\pm 1.63}$}  
        & \textcolor{gray}{$89.08_{\pm 2.42}$}  
        & \textcolor{gray}{$82.91_{\pm 2.01}$}   
        & \textcolor{gray}{$85.29_{\pm 2.89}$}  
        & \textcolor{gray}{$79.40_{\pm 2.31}$} 
        & \textcolor{gray}{$82.64_{\pm 1.78}$} \\
    \midrule
    MASSL \citep{chen2019multi}       & WSL  & $72.54_{\pm 4.24}$ & $73.13_{\pm 4.83}$   & $73.82_{\pm 3.78}$  &  $75.25_{\pm 4.77}$ & $60.91_{\pm 11.28}$  & $59.24_{\pm 13.86}$  & $49.79_{\pm 12.14}$ & $46.26_{\pm 12.68}$  \\
    CPS \citep{chen2021semi} & WSL    &  $71.38_{\pm 6.51}$ & $72.70_{\pm 6.14}$  & $80.42_{\pm 1.71}$  & $83.93_{\pm 2.14}$  &  $70.60_{\pm 5.20}$ &  $67.38_{\pm 6.44}$ &  $63.57_{\pm 8.39}$ &  $62.82_{\pm 8.29}$ \\   
    CCT \citep{ouali2020semi} & WSL  & $74.03_{\pm 2.57}$  &  $75.91_{\pm 3.65}$ & $77.38_{\pm 1.98}$  & $79.49_{\pm 3.23}$  & $67.89_{\pm 4.51}$  & $70.25_{\pm 5.43}$  & $64.80_{\pm 3.65}$ & $66.39_{\pm 4.14}$ \\
    AGMM \citep{wu2023sparsely} & WSL  & $57.35_{\pm 4.11}$   & $70.25_{\pm 4.50}$   &$70.90_{\pm 3.18}$   & $83.82_{\pm 3.23}$  & $60.54_{\pm 3.71}$  & $79.22_{\pm 4.53}$   &$66.76_{\pm 3.88}$  & $72.87_{\pm 4.14}$ \\
    DBDM \citep{luo2022scribble} & WSL  & $51.02_{\pm 7.60}$   & $60.45_{\pm 6.21}$  & $67.18_{\pm 3.48}$  & $79.92_{\pm 4.79}$  & $60.75_{\pm 5.30}$   &  $74.02_{\pm 5.22}$  & $59.71_{\pm 4.23}$  & $77.15_{\pm 4.80}$ \\
    PVA \citep{zhang2023partial} & WSL  & $61.59_{\pm 7.06}$  & $64.88_{\pm 7.46}$  &  $57.91_{\pm 5.76}$ & $60.28_{\pm 5.47}$ & $61.12_{\pm 10.93}$  & $58.41_{\pm 11.07}$  & $62.46_{\pm 4.46}$  & $66.74_{\pm 4.45}$  \\
    Vessel-captcha \citep{dang2022vessel} & WSL  & $76.88_{\pm 4.29}$ & $77.21_{\pm 5.71}$   & $79.98_{\pm 3.30}$  &  $81.88_{\pm 4.05}$ & $69.63_{\pm 9.22}$  & $68.85_{\pm 10.73}$  & $73.08_{\pm 4.21}$ & $74.75_{\pm 4.24}$  \\

    \midrule
    MAE \citep{he2022masked}        & PFM   & $74.68_{\pm 3.31}$ & $74.83_{\pm 4.41}$  & $72.64_{\pm 3.37}$  & $71.23_{\pm 4.60}$  & $68.81_{\pm 3.61}$  & $64.78_{\pm 4.38}$  & $65.86_{\pm 4.65}$  & $63.94_{\pm 4.67}$    \\
    SimCLR \citep{liu2022deep}        & PFM   &$77.75_{\pm 3.13}$  & $81.21_{\pm 3.79}$  & $80.20_{\pm 1.33}$  & $83.74_{\pm 2.02}$  &$66.91_{\pm 6.19}$   & $70.67_{\pm 8.31}$  & $76.02_{\pm 3.40}$  & $79.51_{\pm 3.09}$    \\
    MOCO \citep{he2020momentum}        & PFM   &$78.81_{\pm 4.73}$ & $81.24_{\pm 4.29}$ & \cellcolor{third} $83.96_{\pm 1.81}$ & \cellcolor{third} $87.50_{\pm 2.54}$  & $72.54_{\pm 5.95}$ & $74.42_{\pm 8.44}$   &  $76.01_{\pm 4.37}$  & $79.99_{\pm 4.21}$     \\
    iBOT \citep{zhou2021ibot}        & PFM   &$75.47_{\pm 3.37}$  &$75.90_{\pm 3.99}$ & $78.81_{\pm 2.12}$ & $80.53_{\pm 3.00}$   &  $74.14_{\pm 4.46}$  & $72.10_{\pm 5.84}$  & $67.59_{\pm 4.56}$  & $67.85_{\pm 4.99}$     \\
    SimSiam \citep{chen2021exploring}        & PFM   & $77.88_{\pm 3.48}$ & $80.14_{\pm 3.96}$  & $83.25_{\pm 1.95}$  & $86.03_{\pm 2.35}$  & \cellcolor{third}$80.22_{\pm 4.11}$  & $82.59_{\pm 5.06}$  &\cellcolor{second} $76.29_{\pm 4.35}$  & \cellcolor{second}$80.00_{\pm 4.12}$   \\
    SAM \citep{kirillov2023segment}         & PFM   &  $75.56_{\pm 2.67}$& $79.89_{\pm 3.23}$  & $83.54_{\pm 1.43}$  & $86.92_{\pm 2.14}$   & $76.02_{\pm 3.78}$  & $78.33_{\pm 4.47}$  &\cellcolor{best}$\textbf{76.49}_{\pm 3.24}$ & \cellcolor{best} $\textbf{80.13}_{\pm 2.59}$     \\
    MedSAM \citep{ma2024segment}        & PFM  &\cellcolor{second}  $79.46_{\pm 3.32}$ & \cellcolor{second} $81.59_{\pm 4.39}$  & \cellcolor{second} $84.40_{\pm 1.41}$  & \cellcolor{second} $87.76_{\pm 2.10}$  &  $80.08_{\pm 3.80}$  & \cellcolor{third} $82.66_{\pm 4.20}$  & $76.03_{\pm 3.28}$  & \cellcolor{third}$79.49_{\pm 3.04}$    \\
    \midrule
    iBOT \citep{zhou2021ibot}        & PFM-V   & $77.29_{\pm 4.86}$  & $77.83_{\pm 5.46}$   & $81.30_{\pm 1.64}$  & $82.73_{\pm 2.39}$   &  $77.22_{\pm 3.92}$  & $76.70_{\pm 5.65}$   & $72.21_{\pm 4.02}$  & $74.75_{\pm 4.04}$     \\
    SimSiam \citep{chen2021exploring}        & PFM-V  & $78.56_{\pm 7.03}$ &   $82.31_{\pm 4.60}$  &   $83.77_{\pm 1.57}$  &  $87.30_{\pm 2.02}$  & \cellcolor{second}$80.47_{\pm 4.98}$  & \cellcolor{second}$83.12_{\pm 6.07}$  & $75.11_{\pm 4.85}$   & $79.77_{\pm 4.25}$    \\
    \midrule
    SOCT \citep{kreitner2024synthetic} & DE  & -  & -  & -  & -  &  -  & -  &  -  &  -  \\ 
YoloCurvSeg\citep{lin2023yolocurvseg} & DE  & \cellcolor{third} $78.79_{\pm 3.12}$ & \cellcolor{third} $81.13_{\pm 4.28}$   & $80.89_{\pm 2.41}$  &  $83.45_{\pm 2.73}$ & $78.67_{\pm 3.36}$  & $81.47_{\pm 4.21}$  & $75.04_{\pm 5.32}$ & $78.37_{\pm 5.21}$  \\
    \textbf{Ours(UniVG)} & DE  & \cellcolor{best} $\textbf{80.32}_{\pm 2.89}$ & $ \cellcolor{best} \textbf{82.73}_{\pm 3.89}$  & \cellcolor{best} $\textbf{85.39}_{\pm 1.37}$ & \cellcolor{best}  $\textbf{88.85}_{\pm 1.97}$ & \cellcolor{best}  $\textbf{82.23}_{\pm 3.89}$& \cellcolor{best}$\textbf{85.74}_{\pm 4.05}$ &  $\cellcolor{third}76.16_{\pm 4.02}$  & $79.46_{\pm 4.19}$  \\
    \midrule

        \multirow{2}{*}{Model} & \multirow{2}{*}{Type}  & \multicolumn{2}{c}{FOS-OCTA500} & \multicolumn{2}{c}{FTS-OCT} &  \multicolumn{2}{c}{FBS-DIAS} & \multicolumn{2}{c}{FBS-DSCA} 
    \\ 
    \cmidrule(lr){3-4} \cmidrule(lr){5-6} \cmidrule(lr){7-8} \cmidrule(lr){9-10}

    & & DSC$\uparrow$ & clDice $\uparrow$ & DSC$\uparrow$ & clDice$\uparrow$ & DSC$\uparrow$ & clDice$\uparrow$ & DSC$\uparrow$ & clDice$\uparrow$  \\
    
    \midrule

 \textcolor{gray}{UNet \citep{ronneberger2015u}} 
        & \textcolor{gray}{FIVE}   & \textcolor{gray}{$77.70_{\pm 5.45}$} & \textcolor{gray}{$84.99_{\pm 5.38}$} & \textcolor{gray}{$69.10_{\pm 12.09}$} & \textcolor{gray}{$73.78_{\pm 12.41}$} & \textcolor{gray}{$64.43_{\pm 7.38}$} & \textcolor{gray}{$56.83_{\pm 8.35}$} & \textcolor{gray}{$72.04_{\pm 10.17}$} & \textcolor{gray}{$63.97_{\pm 11.35}$} \\
        \textcolor{gray}{nnU-Net \citep{isensee2021nnu}} 
        & \textcolor{gray}{FIVE} 
        & \textcolor{gray}{$80.88_{\pm 4.99}$}  
        & \textcolor{gray}{$86.93_{\pm 4.71}$}  
        & \textcolor{gray}{$76.49_{\pm 7.97}$} 
        & \textcolor{gray}{$80.23_{\pm 9.26}$}
        & \textcolor{gray}{$73.00_{\pm 7.82}$}  
        & \textcolor{gray}{$71.85_{\pm 7.49}$}  
        & \textcolor{gray}{$81.07_{\pm 7.16}$} 
        & \textcolor{gray}{$79.54_{\pm 8.21}$}    \\
\textcolor{gray}{Retina-Unet \citep{jaeger2020retin}} & \textcolor{gray}{FIVE}  & \textcolor{gray}{$72.61_{\pm 6.99}$} & \textcolor{gray}{$79.72_{\pm 6.91}$}   & \textcolor{gray}{$65.65_{\pm 11.89}$}  &  \textcolor{gray}{$68.45_{\pm 12.51}$} & \textcolor{gray}{$66.15_{\pm 11.65}$}  & \textcolor{gray}{$61.92_{\pm 12.89}$}  & \textcolor{gray}{$75.71_{\pm 9.32}$} & \textcolor{gray}{$72.04_{\pm 11.45}$}  \\
\textcolor{gray}{FSG-Net \citep{bacchin2023fsg}} & \textcolor{gray}{FIVE}  & \textcolor{gray}{$76.62_{\pm 5.61}$} & \textcolor{gray}{$83.51_{\pm 5.31}$}   & \textcolor{gray}{$72.11_{\pm 9.67}$}  &  \textcolor{gray}{$74.97_{\pm 10.47}$} & \textcolor{gray}{$70.31_{\pm 9.49}$}  & \textcolor{gray}{$66.67_{\pm 10.16}$}  & \textcolor{gray}{$76.16_{\pm 8.66}$} & \textcolor{gray}{$72.38_{\pm 10.25}$}  \\

         \textcolor{gray}{UNet \citep{ronneberger2015u}}
        & \textcolor{gray}{FULL}   & \textcolor{gray}{$83.01_{\pm 4.21}$} & \textcolor{gray}{$88.90_{\pm 4.26}$} & \textcolor{gray}{$79.24_{\pm 6.89}$} & \textcolor{gray}{$83.60_{\pm 7.50}$} & \textcolor{gray}{$74.55_{\pm 6.88}$} & \textcolor{gray}{$69.24_{\pm 7.41}$} & \textcolor{gray}{$82.22_{\pm 6.41}$} & \textcolor{gray}{$78.93_{\pm 7.32}$} \\
       \textcolor{gray}{nnU-Net \citep{isensee2021nnu}} 
        & \textcolor{gray}{FULL}   
        & \textcolor{gray}{$85.77_{\pm 2.89}$}   
        & \textcolor{gray}{$90.43_{\pm 2.81}$} 
        & \textcolor{gray}{$80.17_{\pm 5.91}$}  
        & \textcolor{gray}{$83.87_{\pm 7.15}$}  
        & \textcolor{gray}{$78.17_{\pm 4.27}$}   
        & \textcolor{gray}{$74.92_{\pm 4.69}$}  
        & \textcolor{gray}{$89.15_{\pm 4.08}$} 
        & \textcolor{gray}{$88.87_{\pm 4.79}$} \\
\midrule
    
    MASSL \citep{chen2019multi}       & WSL  & $79.28_{\pm 5.59}$& $85.87_{\pm 5.59}$ & $74.65_{\pm 9.12}$ &$77.90_{\pm 10.23}$ & $69.31_{\pm 8.42}$  & $64.11_{\pm 8.90}$   & $72.28_{\pm 10.24}$  & $68.13_{\pm 10.81}$   \\
    CPS \citep{chen2021semi}
    & WSL    & $79.99_{\pm 5.08}$  & $86.69_{\pm 5.07}$  & $72.39_{\pm 12.17}$  &  $77.52_{\pm 12.09}$ & $68.81_{\pm 9.85}$ & $63.03_{\pm 10.86}$& $74.66_{\pm 9.40}$  & $66.55_{\pm 10.82}$    \\
    CCT \citep{ouali2020semi} & WSL  & $73.26_{\pm 6.01}$& $80.99_{\pm 6.80}$  & $66.48_{\pm 9.86}$  &  $67.75_{\pm 11.28}$  & $52.34_{\pm 8.60}$&$40.97_{\pm 6.63}$ &$57.81_{\pm 10.17}$  & $42.68_{\pm 10.44}$     \\

    AGMM \citep{wu2023sparsely} & WSL  & $57.85_{\pm 5.27}$  & $73.64_{\pm 6.88}$  & $57.49_{\pm 6.86}$  & $71.57_{\pm 10.31}$   & $58.30_{\pm 6.01}$  & $58.30_{\pm 7.07}$  & $65.33_{\pm 7.87}$ & $58.75_{\pm 8.96}$ \\
    DBDM \citep{luo2022scribble} & WSL  & $57.09_{\pm 5.30}$ & $67.65_{\pm 6.31}$ & $47.28_{\pm 5.85}$ & $61.47_{\pm 10.55}$   &$53.46_{\pm 6.92}$   &$51.05_{\pm 9.16}$   &$66.51_{\pm 15.10}$&$61.24_{\pm 15.72}$ \\
    PVA \citep{zhang2023partial} & WSL  & $69.45_{\pm 6.65}$  & $77.04_{\pm 6.76}$  & $62.75_{\pm 9.70}$   & $66.36_{\pm 11.51}$  &$59.37_{\pm 7.98}$& $52.84_{\pm 9.82}$  &$63.52_{\pm 9.30}$& $55.34_{\pm 10.18}$ \\
    Vessel-captcha \citep{dang2022vessel} & WSL  & $76.88_{\pm 4.39}$ & $82.84_{\pm 4.66}$   & $72.16_{\pm 7.23}$  &  $76.81_{\pm 8.45}$ & $69.79_{\pm 8.87}$  & $65.72_{\pm 9.25}$  & $77.32_{\pm 6.55}$ & $76.34_{\pm 8.14}$  \\

    \midrule
    MAE \citep{he2022masked}        & PFM   & $77.33_{\pm 5.26}$  & $83.40_{\pm 5.65}$   & $70.70_{\pm 9.21}$  & $72.73_{\pm 10.93}$  & $67.76_{\pm 9.62}$  & $57.27_{\pm 10.19}$  & $76.50_{\pm 8.71}$  &  $71.35_{\pm 10.51}$   \\
    SimCLR \citep{liu2022deep}        & PFM   & $76.96_{\pm 5.81}$ & $84.61_{\pm 5.79}$  &$74.09_{\pm 9.75}$   & $76.64_{\pm 11.25}$  & $66.96_{\pm 7.91}$  & $62.52_{\pm 10.03}$  & $76.38_{\pm 11.32}$  & $74.05_{\pm 13.16}$    \\
    MOCO \citep{he2020momentum}        & PFM   & $77.82_{\pm 6.14}$ & $84.57_{\pm 6.33}$  & $72.67_{\pm 11.58}$  & $75.91_{\pm 12.83}$  &  $64.43_{\pm 13.21}$  &  $60.39_{\pm 12.78}$  & $69.06_{\pm 12.97}$  & $65.03_{\pm 16.14}$    \\
    iBOT \citep{zhou2021ibot}        & PFM   & $75.97_{\pm 6.34}$  & $83. 57_{\pm 5.90}$   & $69.74_{\pm 9..58}$  &$72.24_{\pm 10.69}$ & $69.92_{\pm 6.22}$ &  $60.12_{\pm 6.99}$  & $70.45_{\pm 14.18}$&$61.83_{\pm 16.34}$   \\
    SimSiam \citep{chen2021exploring}        & PFM   & $79.06_{\pm 4.93}$ & $84.79_{\pm 5.02}$  &   $73.69_{\pm 9.77}$ &$77.68_{\pm 10.66}$   &$68.18_{\pm 8.50}$ &$63.49_{\pm 8.06}$    &$75.88_{\pm 7.80}$   &$75.48_{\pm 8.94}$    \\
    SAM \citep{kirillov2023segment}        & PFM   & $\cellcolor{third}80.28_{\pm 4.78}$ &\cellcolor{third} $86.78_{\pm 4.95}$  &\cellcolor{third} $76.16_{\pm 7.79}$  & \cellcolor{third} $79.91_{\pm 8.80}$  & \cellcolor{third}$71.86_{\pm 6.78}$ &\cellcolor{third} $68.42_{\pm 7.44}$   & $78.32_{\pm 5.67}$  & $75.72_{\pm 8.00}$   \\
    MedSAM \citep{ma2024segment}        & PFM  & \cellcolor{second}$80.68_{\pm 4.85}$ & \cellcolor{second} $86.91_{\pm 4.98}$  & \cellcolor{best} $\textbf{76.67}_{\pm 7.28}$  &\cellcolor{second} $80.23_{\pm 8.40}$  &\cellcolor{second} $75.02_{\pm 4.82}$  & \cellcolor{second} $70.32_{\pm 4.69}$  &\cellcolor{third}  $78.41_{\pm 5.45}$ & \cellcolor{third} $76.26_{\pm 5.88}$   \\
    \midrule
    iBOT \citep{zhou2021ibot}        & PFM-V   &$74.71_{\pm 7.16}$  & $82.19_{\pm 6.81}$  & $71.11_{\pm 10.05}$  & $72.26_{\pm 11.22}$  & $70.53_{\pm 7.26}$  & $63.13_{\pm 8.77}$  & $71.96_{\pm 10.35}$  & $64.33_{\pm 12.73}$    \\
    SimSiam \citep{chen2021exploring}        & PFM-V  & $78.21_{\pm 5.32}$ & $86.14_{\pm 5.62}$  &$74.47_{\pm 8.56}$&$78.95_{\pm 9.79}$& $67.41_{\pm 9.07}$  & $60.41_{\pm 8.72}$  & \cellcolor{second} $79.91_{\pm 7.01}$& \cellcolor{second}  $77.94_{\pm 8.56}$   \\
    \midrule
    SOCT \citep{kreitner2024synthetic} & DE  & $80.79_{\pm 4.64}$  & $86.92_{\pm 5.07}$  & -  & -  &  -  & -  &  -  &  -  \\ 
YoloCurvSeg \citep{lin2023yolocurvseg} & DE  &  $80.05_{\pm 4.93}$ &  $86.07_{\pm 5.02}$   & $74.59_{\pm 8.79}$  &  $78.43_{\pm 9.86}$ & $71.54_{\pm 7.93}$  & $68.75_{\pm 8.45}$  & $80.06_{\pm 7.56}$ & $78.83_{\pm 7.71}$  \\

    \textbf{Ours(UniVG)} & DE  &  $\cellcolor{best}\textbf{81.71}_{\pm 4.53}$ &  \cellcolor{best}$\textbf{87.79}_{\pm 4.68}$ & $\cellcolor{second} 76.57_{\pm 7.37}$   & \cellcolor{best} $\textbf{81.02}_{\pm 8.30}$  & $\cellcolor{best}\textbf{75.99}_{\pm 6.95}$ & $\cellcolor{best}\textbf{73.17}_{\pm 7.49}$  & \cellcolor{best} $\textbf{81.11}_{\pm 7.05}$  &  \cellcolor{best} $\textbf{79.86}_{\pm 8.09}$  \\

    \midrule
    \multirow{2}{*}{Model} & \multirow{2}{*}{Type}  &  \multicolumn{2}{c}{FCS-XACD} &  \multicolumn{2}{c}{FCS-SBCD}  &  \multicolumn{2}{c}{FVS-ARIA} & \multicolumn{2}{c}{\textit{Average}} \\ 
    \cmidrule(lr){3-4} \cmidrule(lr){5-6} \cmidrule(lr){7-8} \cmidrule(lr){9-10} 
   
    & & DSC$\uparrow$ &  clDice $\uparrow$ & DSC$\uparrow$ & clDice$\uparrow$ &  DSC$\uparrow$ & clDice$\uparrow$ & DSC$\uparrow$ & clDice$\uparrow$  \\
    
    \midrule
    \textcolor{gray}{UNet \citep{ronneberger2015u}} 
        & \textcolor{gray}{FIVE} 
        & \textcolor{gray}{$66.74_{\pm 7.63}$}  
        & \textcolor{gray}{$69.50_{\pm 7.91}$} 
        & \textcolor{gray}{$72.26_{\pm 6.47}$} 
        & \textcolor{gray}{$58.81_{\pm 8.13}$} 
        & \textcolor{gray}{$62.22_{\pm 10.32}$}  
        & \textcolor{gray}{$66.66_{\pm 10.67}$}  
        & \textcolor{gray}{$66.59_{\pm 7.52}$} 
        & \textcolor{gray}{$64.88_{\pm 10.57}$}  \\

       \textcolor{gray}{nnU-Net \citep{isensee2021nnu}} 
        & \textcolor{gray}{FIVE} 
        & \textcolor{gray}{$72.92_{\pm 7.26}$}  
        & \textcolor{gray}{$75.71_{\pm 7.92}$}  
        & \textcolor{gray}{$80.68_{\pm 4.90}$} 
        & \textcolor{gray}{$71.57_{\pm 6.78}$}
        & \textcolor{gray}{$72.12_{\pm 6.68}$}  
        & \textcolor{gray}{$77.58_{\pm 6.86}$}  
        & \textcolor{gray}{$77.88_{\pm 3.95}$} 
        & \textcolor{gray}{$79.48_{\pm 5.05}$}    \\

\textcolor{gray}{Retina-Unet \citep{jaeger2020retin}} & \textcolor{gray}{FIVE}  & \textcolor{gray}{$57.95_{\pm 8.15}$} & \textcolor{gray}{$58.57_{\pm 8.21}$}   & \textcolor{gray}{$57.31_{\pm 7.95}$}  &  \textcolor{gray}{$51.17_{\pm 7.31}$} & \textcolor{gray}{$62.31_{\pm 10.57}$}  & \textcolor{gray}{$66.23_{\pm 10.88}$}  & \textcolor{gray}{$67.01_{\pm 5.95}$} & \textcolor{gray}{$67.58_{\pm 7.71}$}  \\
\textcolor{gray}{FSG-Net \citep{bacchin2023fsg}} & \textcolor{gray}{FIVE}  & \textcolor{gray}{$62.14_{\pm 9.13}$} & \textcolor{gray}{$61.44_{\pm 10.67}$}   & \textcolor{gray}{$74.30_{\pm 4.63}$}  &  \textcolor{gray}{$64.47_{\pm 5.73}$} & \textcolor{gray}{$64.25_{\pm 9.48}$}  & \textcolor{gray}{$68.98_{\pm 9.77}$}  & \textcolor{gray}{$72.56_{\pm 5.16}$} & \textcolor{gray}{$72.85_{\pm 6.74}$}
\\
 \textcolor{gray}{UNet \citep{ronneberger2015u}}    & \textcolor{gray}{FULL} 
        & \textcolor{gray}{$80.12_{\pm 4.77}$}  
        & \textcolor{gray}{$84.56_{\pm 5.54}$} 
        & \textcolor{gray}{$85.35_{\pm 4.49}$} 
        & \textcolor{gray}{$83.91_{\pm 7.81}$}  
        & \textcolor{gray}{$74.43_{\pm 4.42}$}  
        & \textcolor{gray}{$80.18_{\pm 4.53}$} 
        & \textcolor{gray}{$78.86_{\pm 3.74}$} 
        & \textcolor{gray}{$80.55_{\pm 5.37}$}     \\
         \textcolor{gray}{nnU-Net \citep{isensee2021nnu}}
        & \textcolor{gray}{FULL}   
        & \textcolor{gray}{$84.39_{\pm 3.53}$}   
        & \textcolor{gray}{$89.54_{\pm 3.64}$} 
        & \textcolor{gray}{$86.25_{\pm 3.70}$}  
        & \textcolor{gray}{$78.51_{\pm 6.66}$}  
        & \textcolor{gray}{$89.10_{\pm 10.36}$}   
        & \textcolor{gray}{$92.61_{\pm 9.45}$}  
        & \textcolor{gray}{$83.83_{\pm 3.63}$} 
        & \textcolor{gray}{$85.28_{\pm 5.19}$} \\
    \midrule
    MASSL \citep{chen2019multi}       & WSL  & $70.34_{\pm 7.07}$ & $73.09_{\pm 8.38}$  & $78.86_{\pm 5.05}$  & $68.82_{\pm 7.29}$     &  $56.01_{\pm 12.38}$ &  $59.00_{\pm 13.62}$ &  $68.89_{\pm 8.98}$ & $68.25_{\pm 10.30}$  \\
    CPS \citep{chen2021semi}
    & WSL    & $70.83_{\pm 7.50}$  & $73.03_{\pm 8.05}$  & $79.30_{\pm 5.18}$  &  $67.15_{\pm 6.90}$ &  $66.67_{\pm 9.71}$  & $72.06_{\pm 6.69}$   & $72.60_{\pm 5.26}$ & $72.08_{\pm 7.56}$ \\
    CCT \citep{ouali2020semi} & WSL  & $72.33_{\pm 7.29}$  & $76.33_{\pm 8.42}$   & $75.43_{\pm 5.01}$  &  $62.51_{\pm 6.68}$ & $67.26_{\pm 6.93}$ &  $72.58_{\pm 7.26}$ & $68.09_{\pm 7.31}$ 
 & $66.90_{\pm 12.97}$\\
    AGMM\citep{wu2023sparsely} & WSL  & $64.24_{\pm 7.70}$  & $62.39_{\pm 10.92}$  & $65.41_{\pm 5.89}$  &  $59.26_{\pm 7.74}$ &$64.37_{\pm 6.87}$   &   $72.32_{\pm 7.70}$ & $62.59_{\pm 4.33}$ & $69.31_{\pm 8.18}$ \\
    DBDM \citep{luo2022scribble} & WSL  &$63.31_{\pm 8.31}$   & $65.04_{\pm 10.57}$  &  $57.69_{\pm 8.22}$ &   $54.60_{\pm 9.60}$  &  $58.13_{\pm 5.88}$ & $69.92_{\pm 7.59}$ &  $58.38_{\pm 5.87}$ & $65.68_{\pm 8.67}$ \\
    PVA\citep{zhang2023partial} & WSL  & $59.78_{\pm 8.30}$  &  $60.29_{\pm 9.87}$  & $66.23_{\pm 5.56}$&  $58.42_{\pm 6.57}$& $57.70_{\pm 11.40}$  &  $62.25_{\pm 11.99}$ &  $61.99_{\pm 3.36}$ &  $62.08_{\pm 6.29}$\\
    Vessel-captcha \citep{dang2022vessel} & WSL  & $71.05_{\pm 8.18}$ & $72.56_{\pm 9.70}$   & $68.26_{\pm 10.27}$  &  $58.88_{\pm 8.14}$ & $66.51_{\pm 9.98}$  & $72.58_{\pm 11.08}$  & $72.87_{\pm 4.13}$ & $73.49_{\pm 6.67}$  \\
    \midrule

    \end{tabular}
    \end{adjustbox}
    \subcaption{Pre-lay}
    \end{subtable}
  
\end{table*}
\begin{table*}\ContinuedFloat
\centering
    \begin{subtable}{\textwidth}
    \centering
    \subcaption{Post-lay}
    \begin{adjustbox}{width=\textwidth, totalheight=0.17\textheight}
    \begin{tabular}{llcccccccc}
    \midrule
        MAE \citep{he2022masked}        & PFM   &$69.03_{\pm 6.64}$  &$70.59_{\pm 7.71}$   &$75.52_{\pm 4.66}$   &  $65.44_{\pm 6.38}$  &  $67.95_{\pm 6.35}$ &  $71.91_{\pm 7.55}$  & $71.53_{\pm 3.80}$ & $69.77_{\pm 6.49}$ \\
    SimCLR \citep{liu2022deep}        & PFM   &$74.40_{\pm 6.30}$  & $77.80_{\pm 7.11}$  & $79.52_{\pm 6.70}$  &  $70.24_{\pm 7.70}$  &  $68.33_{\pm 8.29}$ &  $74.83_{\pm 8.38}$  &  $74.32_{\pm 4.60}$ &  $75.98_{\pm 6.22}$ \\
    MOCO \citep{he2020momentum}        & PFM   & $71.73_{\pm 8.99}$ & $74.84_{\pm 10.30}$  &  $80.12_{\pm 4.94}$ &  $70.39_{\pm 6.89}$  &     $70.12_{\pm 7.47}$ &  $76.17_{\pm 7.45}$  & $74.30_{\pm 5.37}$ & $75.50_{\pm 7.66}$\\
    iBOT \citep{zhou2021ibot}         & PFM   &  $70.91_{\pm 8.56}$ & $70.73_{\pm 8.77}$  & $75.68_{\pm 4.94}$  &  $58.64_{\pm 6.87}$  & $67.80_{\pm 6.07}$& $70.11_{\pm 6.09}$   &  $72.41_{\pm 3.58}$ &  $70.33_{\pm 7.61}$   \\
    SimSiam \citep{chen2021exploring}        & PFM   & $69.97_{\pm 8.40}$  & $73.21_{\pm 9.35}$   &  $76.21_{\pm 4.87}$ &    $64.56_{\pm 6.73}$ & \cellcolor{third} $71.33_{\pm 8.38}$  &  \cellcolor{second}$77.57_{\pm 8.41}$ &  $75.63_{\pm 4.33}$ &  $76.87_{\pm 7.05}$   \\
    SAM \citep{kirillov2023segment}        & PFM   & \cellcolor{third}$75.78_{\pm 4.71}$ & \cellcolor{third} $78.28_{\pm 5.64}$ &\cellcolor{third} $80.83_{\pm 4.24}$  & \cellcolor{second} $71.31_{\pm 7.64}$  & $70.12_{\pm 6.20}$   &   $75.01_{\pm 6.47}$   &  $76.81_{\pm 3.68}$ &  $78.24_{\pm 5.38}$  \\
    MedSAM \citep{ma2024segment}        & PFM  & \cellcolor{best}$\textbf{77.84}_{\pm 5.48}$  & \cellcolor{best} $\textbf{81.83}_{\pm 5.87}$  & \cellcolor{second} $81.38_{\pm 4.88}$ &\cellcolor{third} $71.24_{\pm 7.90}$ &  \cellcolor{second}$71.95_{\pm 4.75}$ & \cellcolor{third} $76.93_{\pm 5.07}$ & \cellcolor{second} $78.36_{\pm 3.24}$ & \cellcolor{second} $79.57_{\pm 5.33}$  \\
    \midrule
    iBOT \citep{zhou2021ibot}        & PFM-V   &  $71.48_{\pm 6.78}$ & $69.83_{\pm 7.81}$ & $76.01_{\pm 4.85}$  & $65.65_{\pm 6.45}$  & $69.34_{\pm 6.70}$  &$73.00_{\pm 7.01}$   & $73.92_{\pm 3.50}$   & $73.13_{\pm 6.61}$    \\
    SimSiam \citep{chen2021exploring}        & PFM-V  & $72.39_{\pm 6.83}$ & $76.15_{\pm 7.68}$  & $78.82_{\pm 5.04}$  &   $67.87_{\pm 6.94}$  & $70.16_{\pm 7.26}$ &  $76.10_{\pm 7.16}$ &\cellcolor{third} $77.19_{\pm 3.89}$ &\cellcolor{third} $79.56_{\pm 5.35}$   \\  
    \midrule
    SOCT \citep{kreitner2024synthetic} & DE  & -  & -  & -  & -  &  -  & -  &  -  &  -  \\ 
YoloCurvSeg \citep{lin2023yolocurvseg} & DE  & $74.74_{\pm 6.27}$ & $77.96_{\pm 7.06}$   & $81.06_{\pm 4.97}$  &  $70.94_{\pm 7.83}$ & $71.40_{\pm 8.22}$  & $77.89_{\pm 8.17}$  & $76.98_{\pm 3.47}$ & $78.48_{\pm 4.77}$  \\
    \textbf{Ours(UniVG)} & DE  &\cellcolor{second} $77.39_{\pm 6.00}$ & \cellcolor{second}$81.23_{\pm 6.90}$  & \cellcolor{best}$\textbf{82.74}_{\pm 4.36}$  & \cellcolor{best}$\textbf{73.13}_{\pm 7.00}$ &   \cellcolor{best}$\textbf{73.13}_{\pm 5.45}$ &  \cellcolor{best}$\textbf{78.80}_{\pm 5.51}$ & $\cellcolor{best}\textbf{79.34}_{\pm 3.54}$ &  $\cellcolor{best}\textbf{81.07}_{\pm 4.91}$ \\
    \bottomrule
    \end{tabular}
    \end{adjustbox}
    \end{subtable}
\end{table*}

\begin{table}[h!]

\centering
\caption{Ablation studies on few-shot fundus vascular segmentation and coronary artery segmentation tasks. The contributions of our innovations in combined learning (CL) and few-shot adaptation (FS) are demonstrated.}
\label{tab:ablation_results}
\resizebox{\linewidth}{!}
{
\begin{tabular}{lccccccccccc}
\toprule
\multirow{2}{*}{Task}
& \multicolumn{2}{c}{CL} 
& \multicolumn{2}{c}{FS} 
& \multirow{2}{*}{DSC}
& \multirow{2}{*}{clDice} 
\\
\cmidrule(lr){2-3} \cmidrule(lr){4-5} 
& mask 
& image 
& mask 
& image  
\\
\midrule
\multirow{7}{*}{FVS-CHASEDB1} 
& 
&  
& 
& 
& $72.62_{\pm 3.04}$  
& $73.53_{\pm 3.75}$ 
\\
& $\checkmark$ 
&  
&  
&  
& $82.14_{\pm 1.36}$ 
& $85.62_{\pm 2.32}$ 
\\
& $\checkmark$ 
&  
& $\checkmark$ 
&  
& $83.49_{\pm 1.85}$ 
& $86.77_{\pm 2.70}$ 
\\
&  
&  
&  
& $\checkmark$
& $84.28_{\pm 1.56}$ 
& $87.54_{\pm 2.16}$ 
\\
&  
& $\checkmark$ 
&  
& $\checkmark$ 
& $84.85_{\pm 1.51}$ 
& $87.73_{\pm 2.10}$ 
\\
& $\checkmark$ 
& $\checkmark$ 
&  
& $\checkmark$ 
& $85.04_{\pm 1.39}$ 
& $88.42_{\pm 2.00}$   
\\
& $\checkmark$ 
& $\checkmark$ 
& $\checkmark$ 
& $\checkmark$ 
& $\textbf{85.39}_{\pm 1.37}$ 
& $\textbf{88.85}_{\pm 1.97}$  
\\
\midrule
\multirow{7}{*}{FCS-SBCD} 
&  
&  
&  
&  
& $72.26_{\pm 6.47}$  
& $58.81_{\pm 8.13}$  
\\
& $\checkmark$ 
&  
& 
& 
& $78.58_{\pm 5.13}$ 
& $69.53_{\pm 6.84}$     
\\
& $\checkmark$ 
& 
& $\checkmark$
& 
& $81.12 _{\pm 4.74}$ 
& $69.75_{\pm 6.85}$
\\
& 
& 
& 
& $\checkmark$
& $81.28_{\pm 2.06}$ 
& $71.18_{\pm 7.01}$ 
\\
& 
& $\checkmark$ 
&  
& $\checkmark$
&  $81.98_{\pm 4.68}$ 
& $72.04_{\pm 6.99}$ 
\\
& $\checkmark$ 
& $\checkmark$ 
&  
& $\checkmark$ 
&  $81.26_{\pm 5.03}$ 
&  $69.98_{\pm 7.66}$  
\\
& $\checkmark$ 
& $\checkmark$ 
& $\checkmark$
& $\checkmark$
& $\textbf{82.74}_{\pm 4.36}$ 
& $\textbf{73.13}_{\pm 7.00}$  
\\
\bottomrule
\end{tabular}

}
\end{table}

\subsection{Quantitative evaluation to show metric superiority}
Data-engine-based methods demonstrate significant advantages in few-shot medical image segmentation tasks. Five interesting observations can be made from Tab. \ref{tab: comparative experiment}:
\textbf{1)} Direct learning segmentation methods demonstrate significant performance degradation under few-shot constraints. From the average results, nnU-Net achieves 77.88\% DSC score under the FIVE setting, substantially outperforming specialized vessel segmentation methods Retina-Unet (67.01\%) and FSG-Net (72.56\%). The performance gap is particularly pronounced in few-shot scenarios, with Retina-Unet and FSG-Net performing 10.87\% and 5.32\% lower than nnU-Net, respectively. Although all methods show improvements under the FULL setting, with nnU-Net reaching 83.83\% and UNet achieving 78.86\%, these results confirm that the dependence of direct learning segmentation methods on large-scale annotations fundamentally limits their effectiveness in few-shot learning. 
\textbf{2)} WSL methods are not always superior to supervised learning methods: For example, across all 11 datasets, the average DSC of MASSL, CPS, and CCT is higher than that of UNet, with CPS improving DSC by 6.01\% compared to UNet. This is because unlabeled data contains rich latent information, and semi-supervised learning can leverage this information to enhance the model's generalization and performance. However, on the FBS-DIAS dataset, CCT’s DSC is 12.09\% lower than that of UNet due to the extremely limited labels providing minimal supervision, making the supervision information in semi-supervised settings highly unreliable. 
\textbf{3)} PFM methods show general improvements over WSL and supervised learning methods: Due to the powerful representation capabilities of deep learning, these models achieve comprehensive improvements over WSL and supervised methods. The average DSC of these methods exceeds 70\% across all 11 datasets.
\textbf{4)} PFM-P methods show significant improvement compared to those pre-trained on natural images: Because they learn the feature distribution of vascular images, PFM-P shows notable improvements compared to models pre-trained on natural images. For instance, after being pre-trained on vascular images, iBOT achieves an average DSC improvement of 1.66\% when transferred to downstream vascular segmentation tasks compared to models pre-trained on natural images. Similarly, SimSiam pre-trained on the FCS-XACD dataset using vascular images shows a 2.42\% higher DSC when transferred to downstream tasks compared to the original foundation model pre-trained on natural images.
5) DE methods demonstrate competitive performance but with limited cross-modal adaptability: YoloCurvSeg achieves an average DSC of 76.98\% across all 11 datasets, showing reasonable performance in skeleton-based synthetic data generation. However, our UniVG framework outperforms YoloCurvSeg with an average DSC of 79.34\%, representing a 2.36\% improvement. The performance gap is particularly evident on datasets requiring complex morphological variations, such as FVS-CHASEDB1 where our method achieves 85.39\% DSC compared to YoloCurvSeg's 80.89\%. These results indicate that while algorithmic-based generation methods can provide effective data augmentation, learning-based compositional approaches offer superior adaptability to diverse vascular morphologies and cross-modal scenarios without requiring specialized parameter tuning for each imaging modality. Our UniVG method achieved improvements of 0.92\% Dice and 0.87\% clDice compared to the specialized OCTA generation method SOCT \citep{kreitner2024synthetic} in OCTA vascular segmentation tasks, while the SOCT method cannot be applied to vascular images of other modalities.

Compared to other methods, the DE paradigm demonstrates significant superiority in few-shot medical image segmentation tasks due to its ability to leverage rich latent information from unlabeled data and generate diverse images for training. Our UniVG exhibits outstanding segmentation capabilities across all 11 datasets. In the 5-shot scenario, due to the diversity and authenticity of the generated images, UniVG achieves the highest average DSC (79.34\%) and clDice (81.07\%) across all 11 tasks. MedSAM performs similarly to our framework in terms of average metrics but underperforms on the FBS-DSCA dataset (with a 2.7\% lower DSC than ours). This is because MedSAM lacks representations related to brain DSA and struggles to transfer to unseen medical images, thereby reducing segmentation performance.

\subsection{Qualitative evaluation to show visual superiority}
We present the segmentation results of the proposed UniVG method and 22 comparative methods on five vascular modalities in Fig.~\ref{fig:visualization}. Our UniVG method demonstrates exceptional segmentation accuracy and robust noise resistance in small vascular segmentation tasks, with its robustness being particularly outstanding. Specifically, in the example shown in the first row, UniVG successfully segmented the fine coronary artery vascular, whereas MedSAM and CPS failed to fully segment some of the tiny vascular structures. Although SimCLR was able to segment part of the vascular, its results exhibited noticeable omissions. The SAM method, while providing reasonable segmentation of small vascular, simultaneously introduced false segmentations of non-vascular background regions, resulting in significantly lower segmentation accuracy compared to the UniVG method. Similarly, in other instances, the UniVG method consistently demonstrated higher accuracy and fewer errors in small vascular segmentation tasks. For example, in the results shown in the third row, UniVG successfully segmented the complete fine vascular structure on the left side, while the other comparative methods achieved only partial segmentation. These results fully validate the robustness and generalization capability of the proposed UniVG method in handling complex vascular structures.

\subsection{Ablation studies show improvements of innovations}
\subsubsection{Model ablation study}

To evaluate the effectiveness of our proposed combined learning and few-shot fine-tuning strategies in fundus vascular segmentation and coronary artery segmentation tasks, the FVS-CHASEDB1 and FCS-SBCD datasets were utilized., we conducted detailed ablation studies under a 5-shot setting. The baseline model is UNet (first row), and in the absence of an image-conditioned generator, we used CycleGAN as the conditional generator.
\begin{figure*}
    \centering
    \begin{subfigure}{\linewidth}
        \includegraphics[width=\textwidth]{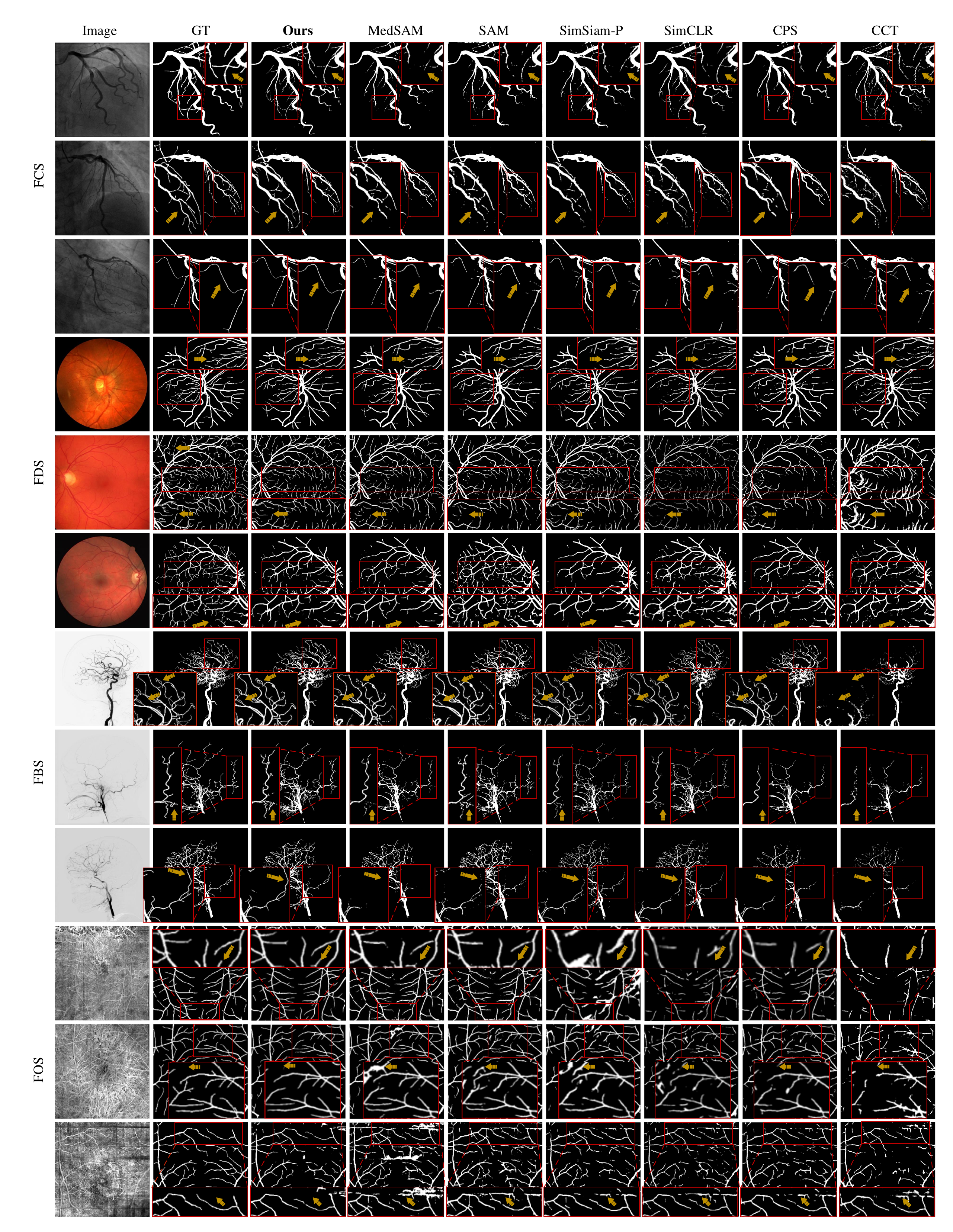}
        \subcaption{Pre-lay}
    \end{subfigure}
    
\end{figure*}
\begin{figure*}\ContinuedFloat
    \centering
    
    \begin{subfigure}{\textwidth}
    \subcaption{Post-lay} 
    \includegraphics[width=\textwidth]{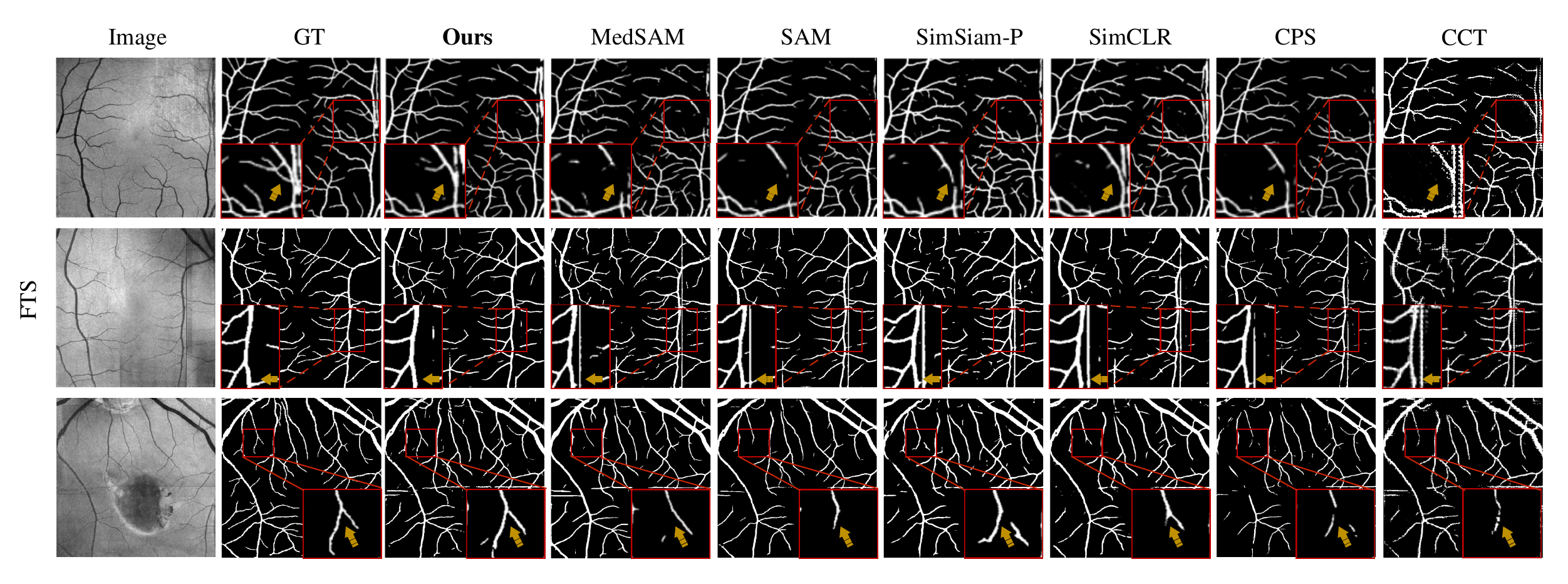}
       
    \end{subfigure}
   
    \caption{Qualitative evaluation shows the visual superiority of our UniVG compared to other benchmark methods.}
    \label{fig:visualization}
    
\end{figure*}

The ablation study results (Tab.\ref{tab:ablation_results}) for the FVS-CHASEDB1 task are shown in the table. The baseline model (without any innovative modules) achieved DSC and clDice scores of 72.62\% and 73.53\%, respectively. Introducing only mask combined learning improved the DSC score by 9.52\%, highlighting the importance of mask-level features in enhancing segmentation performance. Further incorporating mask fine-tuning increased the DSC score to 83.49\%. Using image fine-tuning alone also brought some improvement, with the DSC score reaching 84.28\%. Combining image combined learning with image fine-tuning further improved the DSC score to 84.85\%. Ultimately, applying both mask and image combined learning along with fine-tuning strategies, the best DSC score for the FVS-CHASEDB1  task was 85.39\%, with a clDice score of 88.85\%, demonstrating the effectiveness of all proposed innovations.

In the FCS-SBCD task, the baseline model achieved DSC and clDice scores of 72.26\% and 58.81\%, respectively. Introducing mask combined learning improved the DSC score to 78.58\%. Further incorporating mask fine-tuning enhanced the performance, with the DSC score reaching 81.12\%. Using image fine-tuning alone also led to significant improvements, achieving a DSC score of 81.28\%. Combining Image combined learning with image fine-tuning further boosted the DSC score to 81.98\%. Finally, applying both mask and image combined learning along with fine-tuning strategies, the best DSC score for the FCS-SBCD task was 82.74\%, with a clDice score of 73.13\%, confirming the effectiveness of these innovations in complex medical image segmentation tasks.

\subsubsection{Hyperparameter ablation study}
\begin{figure}
    \centering
    \includegraphics[width=1\linewidth,  keepaspectratio]{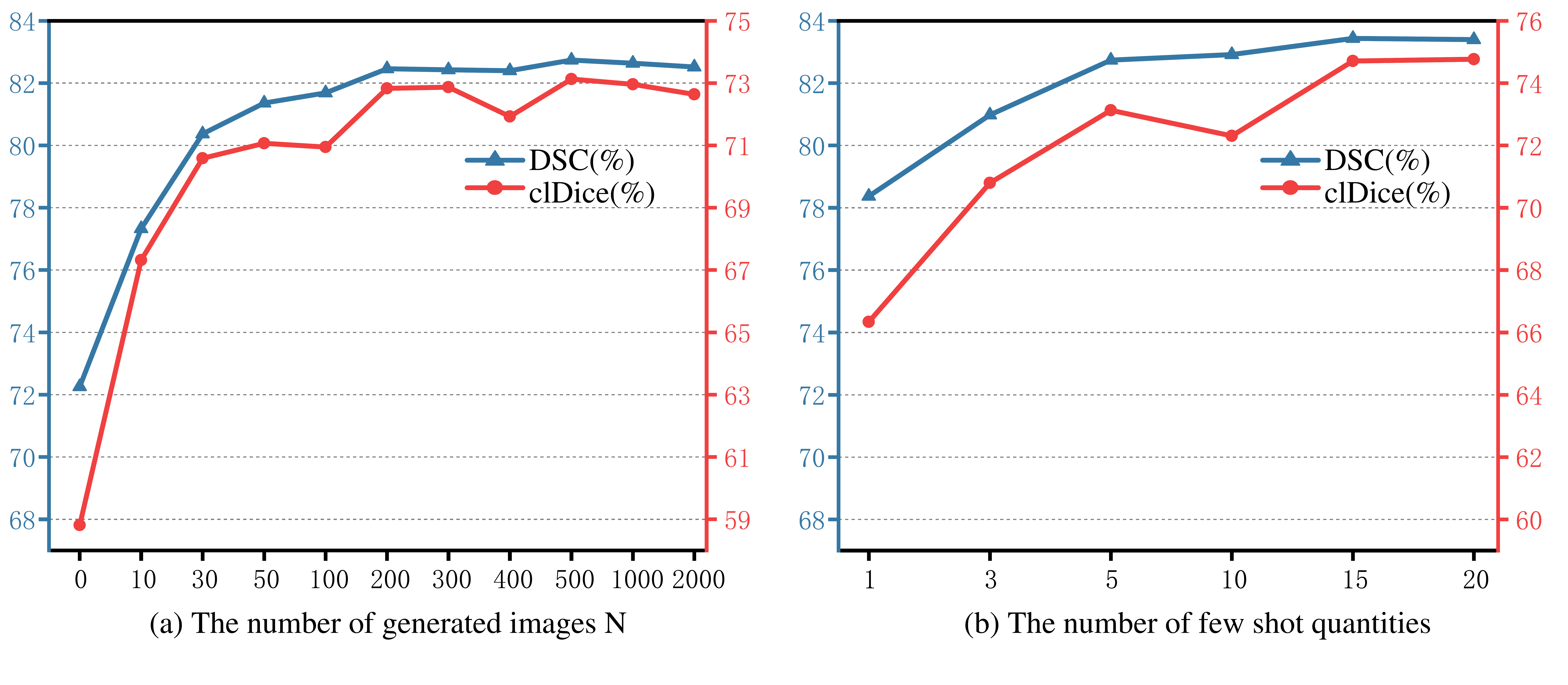}
    \caption{In the few-shot coronary artery segmentation task, (a) analyzes the impact of the hyperparameter related to the number of generated images, while (b) examines the effect of different few-shot quantities on the downstream vascular segmentation performance.}
    \label{fig:DataGenerationAlbation}
\end{figure}

We conducted ablation experiments on the hyperparameter related to the number of generated images in the five-shot coronary artery segmentation task. The results are shown in Fig. \ref{fig:DataGenerationAlbation} (a). From the figure, it can be observed that as the parameter increases, the performance of the segmentation model gradually improves. Specifically, the performance improvement from increasing the number of generated images from 10 to 200 is greater than that from increasing it from 200 to 2000. It is worth noting that when the number of generated images is set to 0, the training process does not involve any generated images.

We also observed the impact of different numbers of few-shot samples on the final segmentation model performance during downstream task adaptation. The results are shown in Fig. \ref{fig:DataGenerationAlbation}   (b). As can be seen from the figure, the performance of the segmentation model gradually improves with an increase in sample quantity. Specifically, the most significant performance enhancement occurs when the sample number increases from 1 to 3. Further increasing the sample size up to 15 leads to performance stabilization.

\subsection{Framework analysis}
\subsubsection{Analysis of mask generation}

\begin{figure}[t!]
    \centering
     \includegraphics[width=\linewidth]{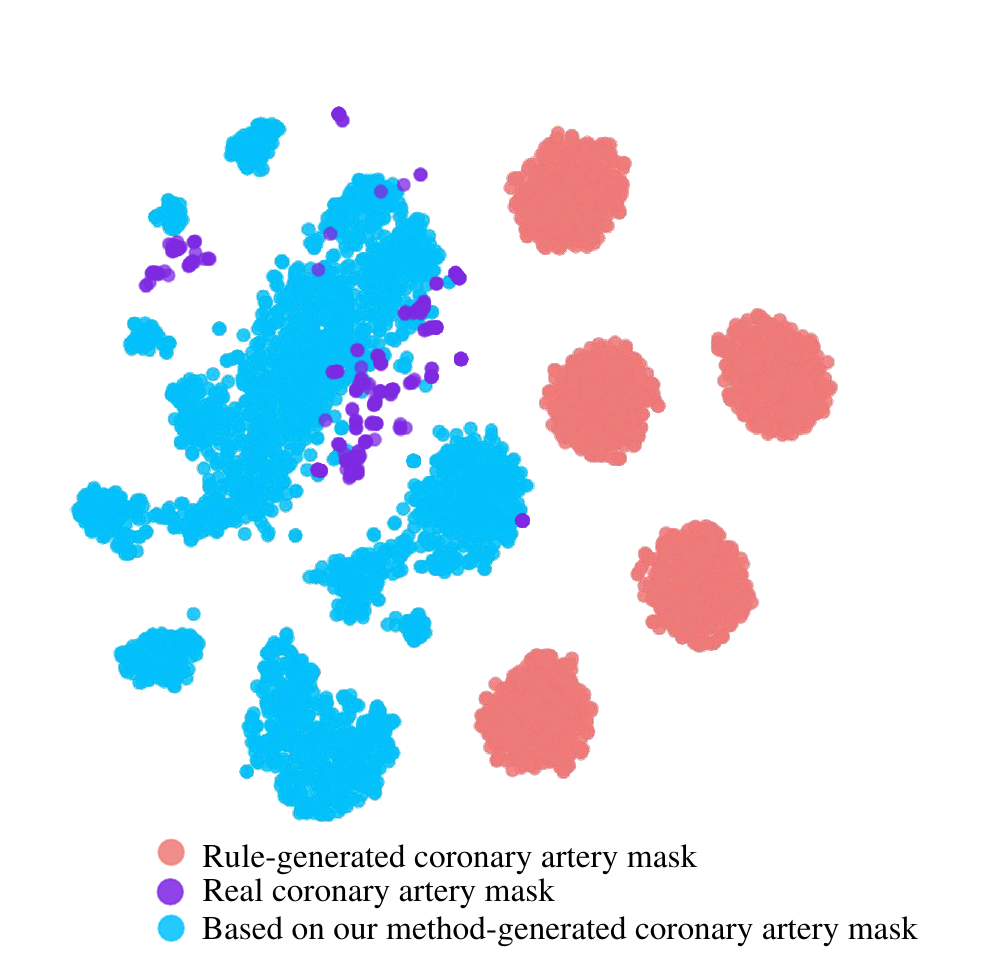}
    \caption{t-SNE visualization comparing the distribution of coronary artery masks generated by our method, rule-generated coronary artery masks, and real coronary artery masks. The masks generated by our method show better uniformity and wider distribution compared to rule-generated masks, indicating high fidelity and diversity.}
    \label{fig:t-SNE}
\end{figure}

\textbf{1) Cluster analysis of generated vascular masks:} The masks generated by our method possess both high fidelity and diversity. We compared the distributions of rule-generated coronary artery masks and masks generated by our method against real coronary artery masks. From the t-SNE visualization \citep{van2008visualizing} in Fig. \ref{fig:t-SNE}, it can be seen that the masks generated by our method are generally consistent with and well mixed with the real masks. Compared to Rule-generated coronary artery masks, the masks generated by our method are more uniformly and widely distributed, indicating that our method produces masks with both high fidelity and diversity.

\begin{figure}[t!]
    \centering
     \includegraphics[width=\linewidth]{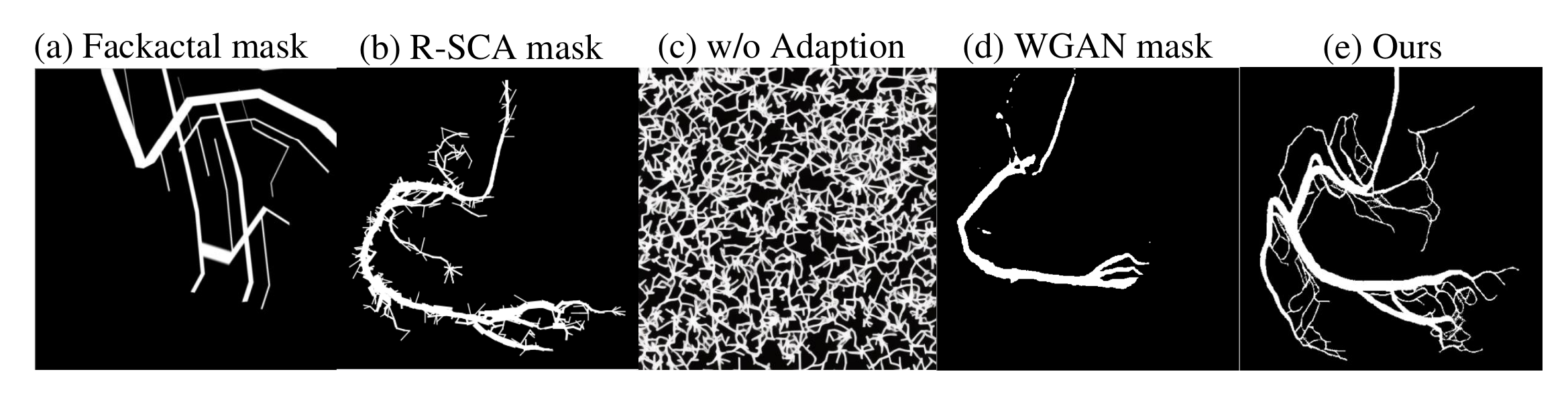}
    \caption{Comparison of coronary artery masks generated by different methods: (a) bifurcation algorithm, (b) R-SCA vascular generation algorithm, (c) vascular mask data-engine without adaption, (d) WGAN, (e) our approach. In comparison to other mask generation methods, the vascular masks generated by our approach exhibit high visual authenticity.}
    \label{fig:MaskImageComparison}
\end{figure}

\textbf{2) Visualization of generated masks with different methods: }The masks generated by our method exhibit high visual authenticity. A more intuitive observation of the advantages of our method can be seen in Fig. \ref{fig:MaskImageComparison}. The mask generation methods from left to right are: (a) bifurcation algorithm \citep{ma2021self}, (b)
R-SCA is an adapted spatial colonization algorithm that incorporates real vessel topology as initialization, 
(c) vascular mask data-engine  without adaption, (d) WGAN \citep{arjovsky2017wasserstein}, and (e) our approach. It can be observed that the masks generated by (a), (b), and (c) differ significantly from the real coronary artery images. Although the overall vascular morphology generated by (d) WGAN is close to real vessels, vessel discontinuities have occurred.  Only (e) our approach produces coronary artery images that are closest to the real ones.

\textbf{3) Vascular generation methods comparison and downstream impact:} Our proposed UniVG method outperforms existing methods in terms of structural authenticity for vascular generation. We investigated the structural authenticity of generated vascular masks by comparing four different generation approaches: fractal mask generation \citep{ma2021self}, SCA \citep{runions2005modeling,runions2007modeling}, WGAN \citep{arjovsky2017wasserstein}, and our proposed UniVG method, as presented in Table \ref{tab:fid_comparison}. The results demonstrate that our UniVG method achieved the highest structural authenticity with the lowest FID score of 80.90. Among rule-based methods, SCA showed moderate performance with an FID score of 147.32, while fractal mask generation exhibited the poorest authenticity with an FID score of 338.54. For deep learning-based approaches, WGAN achieved an intermediate FID score of 231.96, indicating better performance than rule-based methods but inferior to our approach. All methods were evaluated under identical conditions, generating 500 coronary artery masks that were compared against the same 295 real coronary artery images from the SBCD dataset for FID calculation. Our method achieved the highest authenticity (FID score of 80.90) and correspondingly delivered the best segmentation performance with Dice and clDice scores of 82.74\% and 73.13\%, respectively. These experiments demonstrate that the authenticity of generated masks provides guarantees for the accuracy of vascular segmentation.

\begin{table}[h]
\centering
\caption{Comparison of structural authenticity and downstream segmentation performance across different vascular mask generation methods on the SBCD dataset. FID scores are calculated between 500 generated coronary artery masks and 295 real coronary artery masks.}
\label{tab:fid_comparison}
\resizebox{0.48\textwidth}{!}{
\begin{tabular}{lcccc}
\toprule
Method & Category & FID ↓ & Dice ↑ & clDice ↑ \\
\midrule
Fractal mask \citep{ma2021self} & Rule-based & 338.54 & 81.15±4.89 & 70.33±7.49 \\
SCA \citep{runions2005modeling,runions2007modeling} & Rule-based & 147.32 & 81.98±4.68 & 72.04±6.99 \\
WGAN \citep{arjovsky2017wasserstein} & Deep learning-based & 231.96 & 81.74±4.79 & 71.25±7.17 \\
Ours(UniVG) & Deep learning-based & 80.90 & 82.74±4.36 & 73.13±7.00 \\
\bottomrule
\end{tabular}
}
\end{table}

\begin{figure}[t!]
    \centering
     \includegraphics[width=\linewidth]{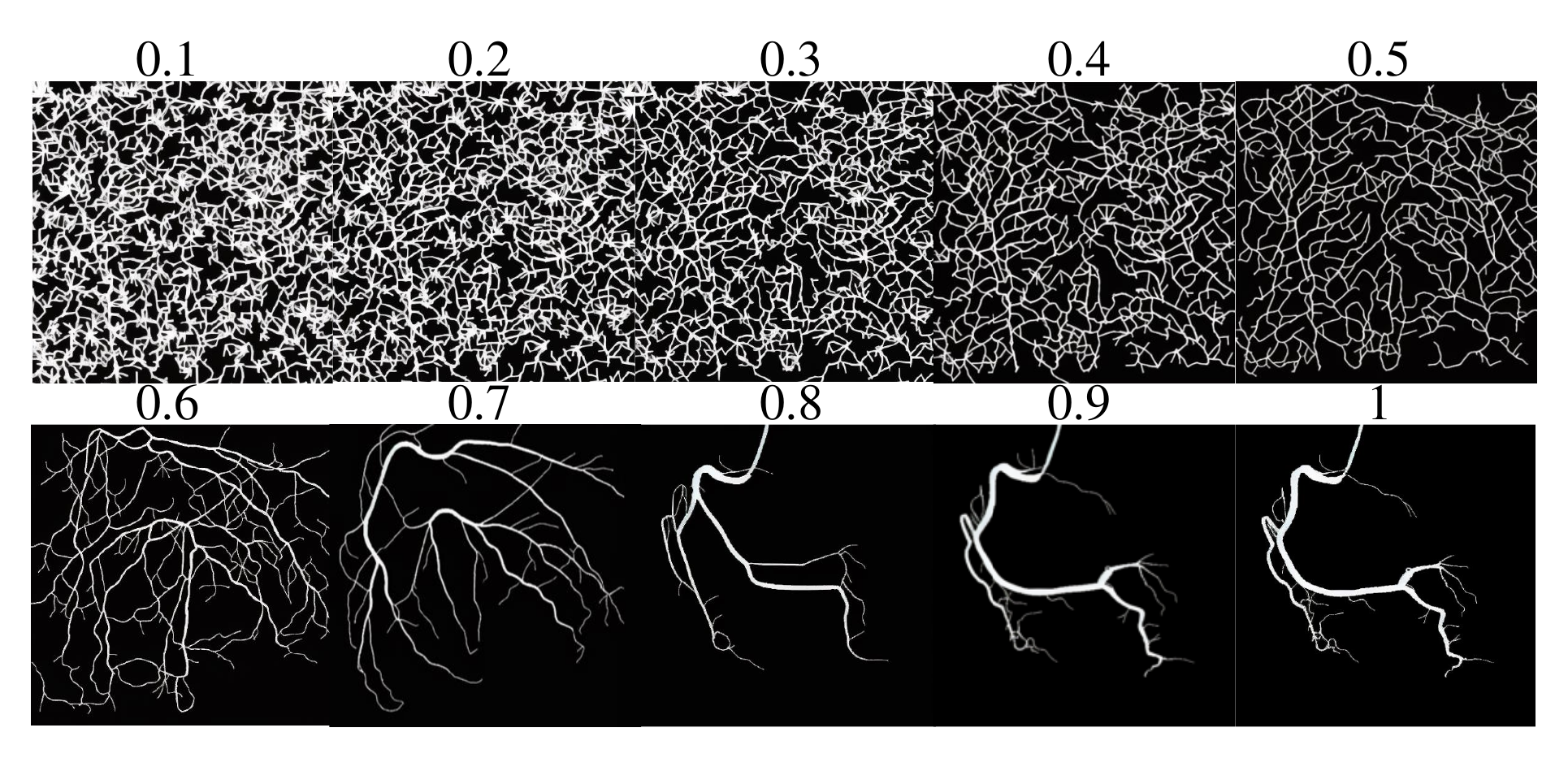}
    \caption{Generation results of coronary artery masks with different values of fusion hyperparameter $\lambda_1$
  during downstream few-shot adaption.}
    \label{fig:LohaHypeParametes}
\end{figure}

\textbf{4)} Fusion parameters affect masks: We analyzed the fusion parameter $\lambda_1$
in Equ. \ref{equation: Hyperparameters} of the Few-shot Generative Adaptation for mask data-engine. After training the model, we assigned different values to $\lambda_1$
and observed its impact on the generation results, as shown in Fig. \ref{fig:LohaHypeParametes}. It can be observed that as  $\lambda_1$
increases up to 1, the generated masks transition from not conforming to the coronary distribution at all to gradually approaching the true coronary distribution.

5) \textbf{LoRA and LoHA combination validation:}  The combination of LoRA and LoHA methods demonstrates superior performance compared to individual fine-tuning strategies in few-shot coronary artery segmentation tasks. To conduct this validation, we employed a few-shot learning approach using five identical masks from the SBCD dataset as fine-tuning samples. The fine-tuned generative model was then used to directly generate 500 synthetic masks, which subsequently served as conditioning inputs to generate 500 corresponding coronary artery images. The generated synthetic data was then combined with the original five real images to train segmentation models for each method comparison. Model performance was evaluated on the SBCD test set to ensure fair comparison across methods. As shown in Table \ref{tab:lora_loha_comparison}, experimental comparison of LoRA-only, LoHA-only, and hybrid LoRA+LoHA approaches demonstrates that the hybrid method achieves superior performance with Dice coefficient of 82.74\% and clDice of 73.13\%. The hybrid approach outperforms individual methods by 1.48\% and 0.99\% in Dice coefficient compared to LoRA-only and LoHA-only respectively, and by 3.12\% and 2.49\% in clDice metric respectively, confirming the synergistic benefits of the combination strategy.
\begin{table}[h]
\centering
\caption{Comparative performance of different fine-tuning strategies for 5-shot coronary artery segmentation on the SBCD dataset.}
\label{tab:lora_loha_comparison}
\begin{tabular}{lcc}
\toprule
Method & DSC ↑ & clDice ↑ \\
\midrule
LoRA-only & $81.26_{\pm4.82}$ & $70.01_{\pm7.33}$ \\
LoHA-only & $81.75_{\pm4.68}$ & $71.64_{\pm7.26}$ \\
LoRA+LoHA & $\textbf{82.74}_{\pm4.36}$ & $\textbf{73.13}_{\pm7.00}$ \\
\bottomrule
\end{tabular}
\end{table}

\subsubsection{Analysis of image generation}
\begin{table}[t]
\centering
\caption{On the SBCD dataset in the few-shot coronary artery segmentation task, the impact of images generated by different conditional generative models on the downstream segmentation results is analyzed.}
\label{tab: different models}
\resizebox{0.48\textwidth}{!}{
\begin{tabular}{cccc}
\toprule
Model & FID ↓
& DSC ↑
& clDice ↑
\\
\midrule

pix2pix \citep{isola2017image}    
& $258.38$ 
& $80.86_{\pm 4.91}$ 
& $70.62_{\pm 7.00}$
\\
CycleGAN \citep{zhu2017unpaired}
& $327.92$ 
& $81.12_{\pm 4.74}$ 
& $69.75_{\pm 6.85}$ 
\\
Our \citep{zhang2023adding} 
& $\textbf{123.04}$
& $\textbf{82.74}_{\pm 4.86} $
& $\textbf{73.13}_{\pm 7.00} $ 
\\
\bottomrule
\end{tabular}}
\label{tab:comparison}
\end{table}

\textbf{1) Effects of different conditional generative models on image quality and segmentation performance:} UniVG method significantly outperforms traditional generative models in both image generation quality and downstream segmentation task performance. We investigated the effectiveness of conditional generative models by evaluating both image authenticity and downstream segmentation performance across three approaches: UniVG, pix2pix \citep{isola2017image}, and CycleGAN \citep{zhu2017unpaired}, as presented in Tab. \ref{tab: different models}. In terms of image authenticity, UniVG achieved the highest quality with an FID score of 123.04, while pix2pix and CycleGAN showed significantly higher FID scores of 258.38 and 327.92, respectively, indicating that compared to the other two methods, UniVG generates images with higher perceptual similarity to real images. This superior image quality directly translated to better segmentation performance, as our proposed UniVG with compositional learning attained the highest downstream performance with a DSC of 82.74\% and a clDice of 73.13\%. In comparison, pix2pix and CycleGAN correspondingly showed lower clDice scores, with reductions of 2.51\% and 3.38\%, respectively. Notably, all models were trained under identical conditions, utilizing the same five labeled coronary artery images to generate 500 synthetic images, with FID scores calculated against 295 real coronary artery images and segmentation performance evaluated using identical conditional masks.

\textbf{2) Effects of different network architectures on segmentation performance:}  We explored the scenario of replacing the downstream segmentation model with alternative architectures, comparing the performance of three distinct models: UNet, VMUNet \citep{ruan2024vm}, and Swin-UNet \citep{cao2022swin}, in downstream segmentation tasks. As shown in Tab. \ref{tab:comparison}, Swin-UNet achieved the highest performance in coronary artery segmentation, with a DSC of 83.17\% and a clDice of 73.24\%. These results demonstrate that higher segmentation performance can be achieved by flexibly substituting the downstream segmentation model, highlighting the adaptability and versatility of our framework.
\begin{table}[h]
\centering
\caption{On the SBCD dataset in the few-shot coronary artery segmentation task, the impact of different downstream segmentation models on the segmentation results is analyzed.}
\begin{tabular}{ccc}
\toprule
Model & DSC & clDice  \\
\midrule
VMUnet \citep{ruan2024vm}        & $80.56_{\pm 4.89} $ & $67.66_{\pm 7.65}$  \\
UNet \citep{ronneberger2015u}           & $82.74_{\pm 4.86} $ & $73.13_{\pm 7.00}$ \\
Swin-Unet \citep{cao2022swin}       & $\textbf{83.17}_{\pm 4.54} $ & $\textbf{73.24}_{\pm 6.85} $  \\
\bottomrule
\end{tabular}
\label{tab:comparison}
\end{table}

\section{DISCUSSION}
\textbf{1) Comparison with SOTA methods on training data requirements and performance}: Our proposed approach demonstrates superior data efficiency while maintaining competitive segmentation performance across diverse vessel imaging modalities. Table \ref{tab:sota_comparison}  demonstrates that our approach achieves competitive performance compared to state-of-the-art methods across 11 vessel segmentation datasets while requiring significantly fewer training samples. Although our method uses only 5 training samples compared to the 15 to 240 samples needed by SOTA methods, it maintains comparable segmentation accuracy on most datasets. Notably, on the CHASEDB1 dataset, our approach (85.39\%) even outperforms the SOTA method (84.01\%). These results indicate that our few-shot learning approach can preserve competitive segmentation precision while dramatically reducing training data requirements, which holds significant practical value for medical image analysis scenarios where annotated data is scarce.

\begin{table}[htbp]
\centering
\caption{Performance comparison between SOTA methods and our approach on 11 vessel segmentation datasets.}
\label{tab:sota_comparison}

\resizebox{0.48\textwidth}{!}{
\begin{tabular}{lcccc}
\toprule
\textbf{Datasets} & \textbf{SOTA} & \textbf{SOTA} & \textbf{Our} & \textbf{Our} \\
\textbf{(SOTA reference)} & \textbf{train images} & \textbf{DSC} & \textbf{train images} & \textbf{DSC} \\
\midrule
DRIVE \citep{bhati2025dynamic} & 20 & 84.68 & 5 & 80.32 \\
CHASEDB1 \citep{bhati2025dynamic} & 20 & 84.01 & 5 & 85.39 \\
HRF \citep{bhati2025dynamic} & 15 & 83.89 & 5 & 82.23 \\
ORVS \citep{sarhan2021transfer} & 42 & 78.11 & 5 & 76.16 \\
OCTA500 \citep{song2024optimized} & 180 & 88.68 & 5 & 81.71 \\
DIAS \citep{liu2024dias} & 30 & 78.22 & 5 & 75.99 \\
DSCA \citep{xie2024dsnet} & 56 & 89.34 & 5 & 81.11 \\
XACD \citep{song2024optimized} & 75 & 83.14 & 5 & 77.39 \\
SBCD \citep{isensee2021nnu} & 240 & 86.25 & 5 & 82.74 \\
ARIA \citep{isensee2021nnu} & 108 & 89.10 & 5 & 73.13 \\
OCT \citep{isensee2021nnu} & 240 & 80.17 & 5 & 76.57 \\
\bottomrule
\end{tabular}
}
\end{table}

\textbf{2) Impact of inadequate pretraining on model performance}: Insufficient generation pretraining causes performance degradation, but the impact remains limited. To evaluate our framework's sensitivity to suboptimal training conditions, we conducted experimental analysis examining how inadequate pretraining steps affect the quality of generated images and subsequent segmentation performance. On the SBCD dataset, we compared three configurations: 1) UNet baseline trained with only 5 labeled coronary artery images, 2) UniVG with insufficient pretraining at 100,000 steps, and 3) UniVG with adequate pretraining at 141,000 steps. Both UniVG configurations used the same 500 coronary artery mask images as conditions to generate corresponding coronary artery images, which were then combined with the 5 real images for training the segmentation model. Table \ref{tab:pretraining_steps} demonstrates that while insufficient pretraining (100,000 steps) causes performance degradation compared to adequate pretraining (141,000 steps), with Dice scores decreasing by 1.28\% and clDice scores declining by 2.29\%, the impact remains manageable. Both UniVG configurations substantially outperformed the UNet baseline, confirming that our framework maintains robust performance levels even under suboptimal training conditions.

\begin{table}[htbp]
\centering
\caption{Impact of pretraining steps on coronary artery segmentation performance: insufficient and adequate training comparison on the SBCD dataset.}
\label{tab:pretraining_steps}

\resizebox{0.48\textwidth}{!}{
\begin{tabular}{lccc}
\toprule
\textbf{Model} & \textbf{Pretraining train steps} & \textbf{DSC$\uparrow$} & \textbf{clDice$\uparrow$} \\
\midrule
UNet \citep{ronneberger2015u} & - & 72.26$\pm$6.47 & 58.81$\pm$8.13 \\
\multirow{2}{*}{Our (UniVG)} & 100,000 & 81.46$\pm$4.82 & 70.84$\pm$7.07 \\
 & 141,000 & \textbf{82.74$\pm$4.36} & \textbf{73.13$\pm$7.00} \\
\bottomrule
\end{tabular}
}
\end{table}

\begin{figure*}[t!]
    \centering
    \includegraphics[width=\textwidth]{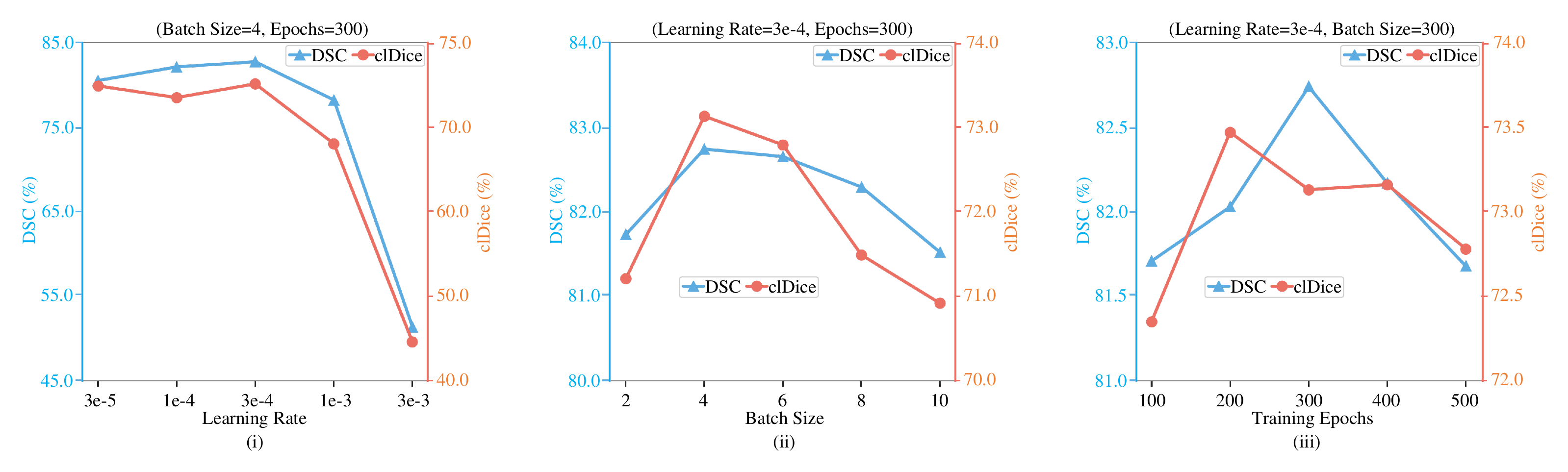}
    \vspace{-1cm}
    \caption{Hyper-parameter sensitivity analysis on SBCD dataset for coronary artery vessel segmentation.}
     \vspace{-0.2cm}
    \label{fig:hyparamter_comparative}
\end{figure*}

\textbf{3) Multidimensional hyperparameter sensitivity evaluation:} We designed 15 different hyperparameter configurations on the SBCD coronary artery dataset to evaluate model robustness by systematically adjusting learning rate (3e-5 to 3e-3), batch size (2 to 10), and training epochs (100 to 500). As shown in Fig.~\ref{fig:hyparamter_comparative}, learning rate sensitivity analysis (Fig.~\ref{fig:hyparamter_comparative}(i)) demonstrates that 3e-4 is the optimal choice (DSC: 82.74\% and clDice: 73.13\%) while excessive learning rate (3e-3) causes dramatic performance degradation to DSC 51.37\%. Batch size sensitivity analysis (Fig.~\ref{fig:hyparamter_comparative}(ii)) reveals the model's good tolerance to batch size variations with DSC fluctuating within 81.52\%-82.74\% range and performance differences of only 1.22\%. Training epochs sensitivity analysis (Fig.~\ref{fig:hyparamter_comparative}(iii)) indicates that 300 epochs achieve optimal performance while both insufficient training (100 epochs) and excessive training (500 epochs) lead to performance degradation. The experiment validates the model's robustness within appropriate hyperparameter ranges with learning rate being the most critical sensitivity factor.

\begin{figure}[htbp]
    \centering
    \resizebox{0.5\textwidth}{!}{  \includegraphics[width=0.5\textwidth, height=0.5\textheight, keepaspectratio]{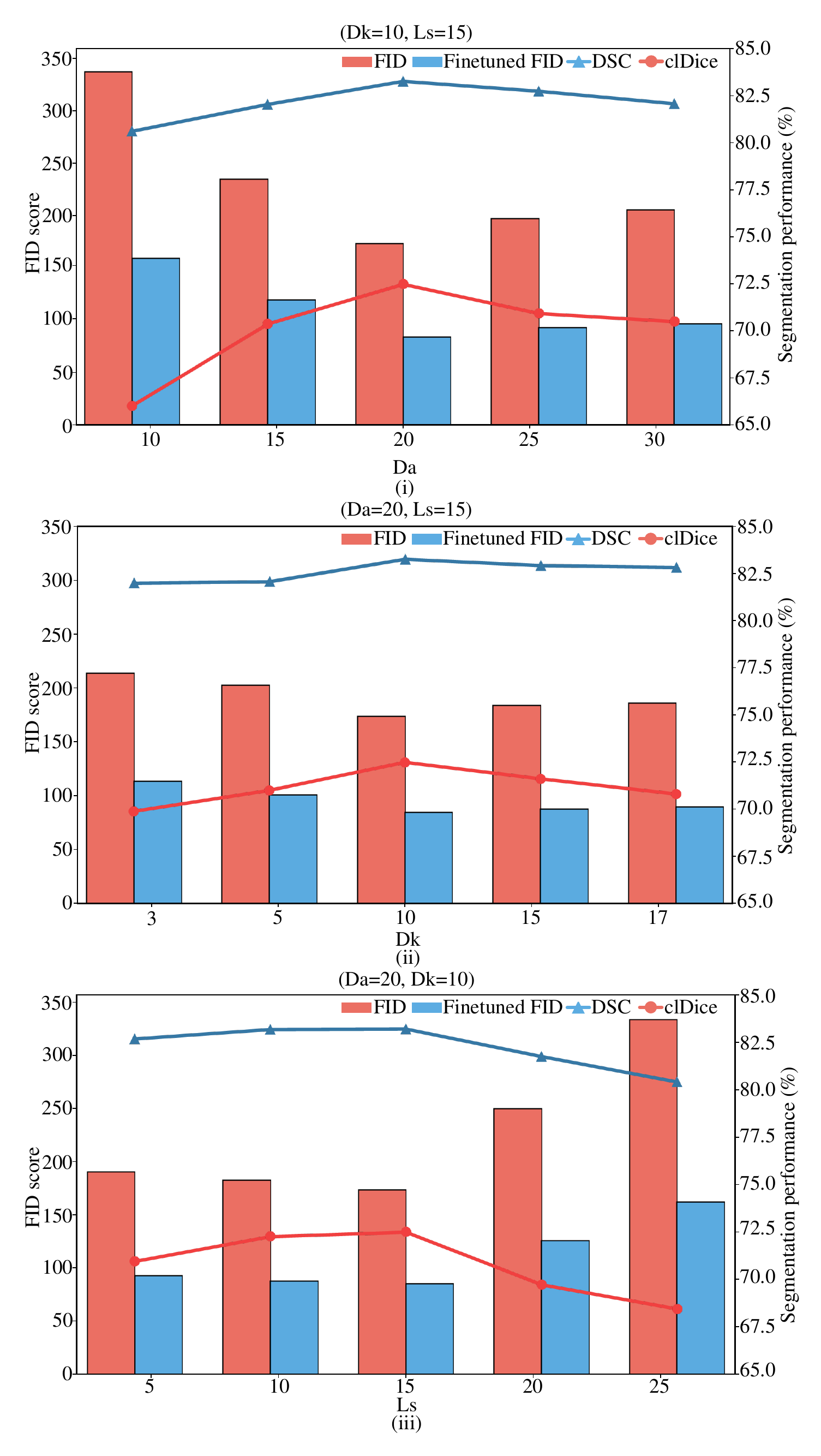}
    }
    \vspace{-1cm}
    \caption{Parameter sensitivity analysis of SCA parameters on generated vascular structure quality. Da represents attraction distance (pixels), Dk represents kill distance (pixels), Ls represents segment length (pixels).}
   
\label{fig:parameter_sensitivity}
\end{figure}

\textbf{4) SCA parameter sensitivity evaluation for vascular generation quality optimization:} To systematically evaluate the influence of SCA parameters on vascular generation quality, we conducted comprehensive sensitivity analysis by varying attraction distance Da= \{10, 15, 20, 25, 30\} pixels, kill distance Dk = \{3, 5, 10, 15, 17\} pixels, and segment length Ls = \{5, 10, 15, 20, 25\} pixels while maintaining other parameters constant. For each configuration, 2000 coronary artery images were generated as pre-training data, with FID evaluation using 500 generated images compared against 295 real SBCD coronary artery masks, and segmentation performance assessed on the SBCD test set. As shown in Fig.\ref{fig:parameter_sensitivity}, the attraction distance analysis Fig.\ref{fig:parameter_sensitivity}.(i) demonstrates performance peaking at Da=20 pixels with optimal FID scores (173.62 direct, 90.43 fine-tuned) and highest segmentation performance (Dice: 82.71±4.21\%, clDice: 72.65±6.59\%), while extreme values significantly degrade quality. The kill distance analysis Fig.\ref{fig:parameter_sensitivity}.(ii) shows Dk=10 achieves optimal performance, with both lower (Dk=3, FID: 211.31) and higher values (Dk=17, FID: 185.30) demonstrating reduced quality. The segment length analysis Fig.\ref{fig:parameter_sensitivity}.(iii) reveals Ls=15 achieves best performance, with extreme values like Ls=25 showing dramatically increased FID (324.41), providing practical guidance for parameter tuning across different application scenarios.

\textbf{5) Biophysical plausibility validation through Murray's law compliance analysis:} Our experimental validation demonstrates that our vascular generation method can achieve equivalent biophysical properties to real vascular networks. We conducted Murray's law\citep{murray1926physiological} compliance analysis on both the real vascular masks from the DRIVE dataset and 500 generated vascular mask images. The evaluation framework analyzed vessel bifurcation points and computed diameter relationships according to Murray's law: $r_0^{\gamma} = r_1^{\gamma} + r_2^{\gamma}$, where $\gamma$ represents the branching exponent and $r_0$, $r_1$, $r_2$ denote parent and child vessel radii respectively. For each bifurcation point, the Murray ratio (MR) was calculated as $MR = \sum r_i^{\gamma}/r_0^{\gamma}$, with deviation quantified as $|1 - MR|$. The compliance score was computed using $CS = 1/{(1 + \text{mean\_deviation})}$, where higher scores indicate better adherence to Murray's law. We processed 40 real retinal vessel masks from the DRIVE dataset and 500 generated masks using our UniVG framework, with $\gamma$ set to 2 \citep{jessen2023branching}. As shown in Table~\ref{tab:murray_law}, our generated vessels achieved mean compliance scores of 74.61\%, compared to 74.74\% for real vessels, representing a difference of only 0.13\%. These results validate that our vascular generation method has preserved the fundamental biophysical characteristics essential for realistic vascular network generation.
\begin{table}[htbp]
\centering
\caption{Murray's law compliance comparison between real and generated vascular masks. Compliance scores (CS) indicate biophysical plausibility, with higher scores indicating better adherence to Murray's law.}
\label{tab:murray_law}

\begin{tabular}{lcc}
\toprule
\textbf{Mask category} & \textbf{Real mask} & \textbf{Generated mask} \\
\midrule
\textbf{CS} & 74.74$\pm$2.0 & 74.61$\pm$4.2 \\
\bottomrule
\end{tabular}
\end{table}

\begin{figure}[h]
\centering
\includegraphics[width=0.5\textwidth]{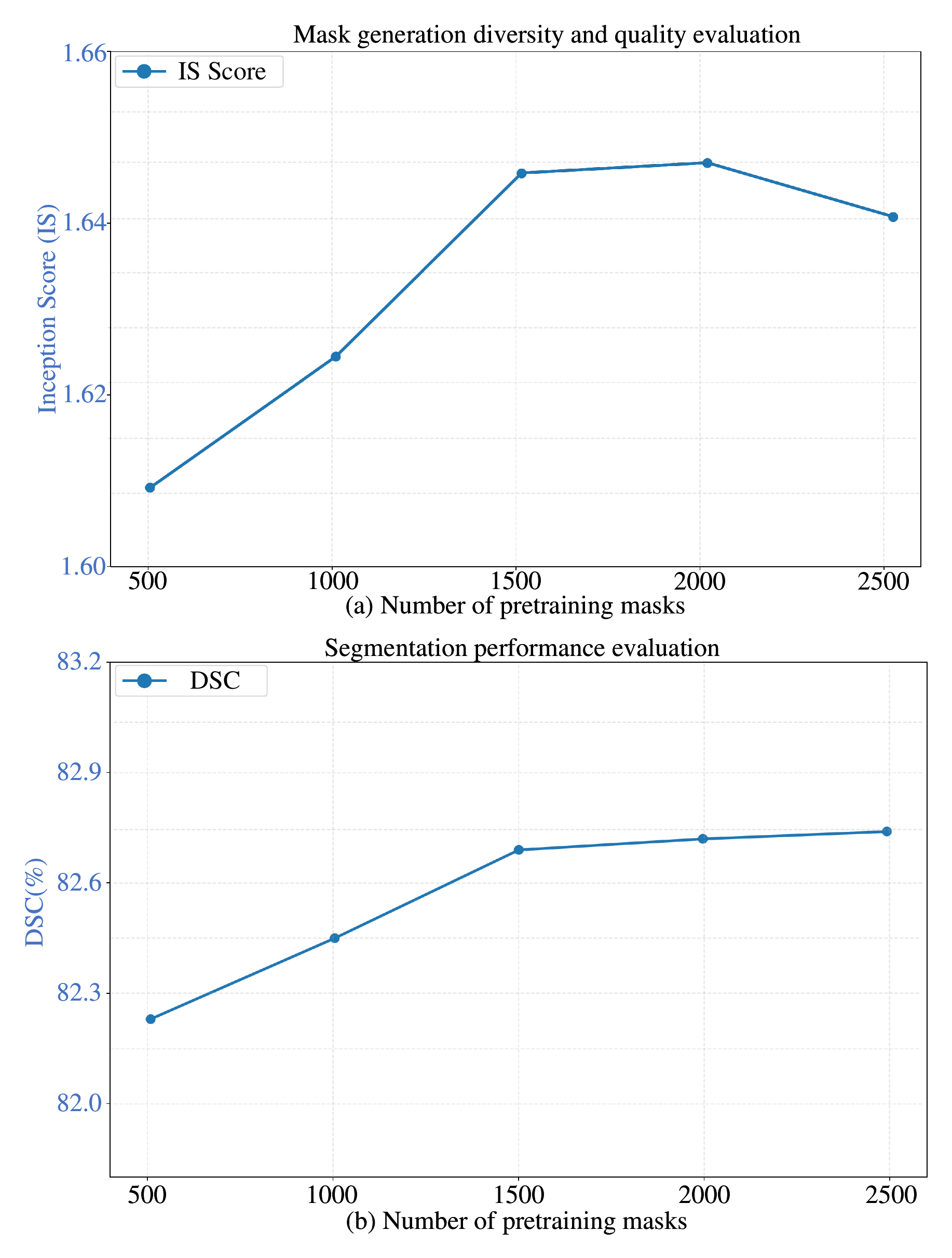}
\caption{ Performance of synthetic mask diversity and downstream segmentation across different scales of pretraining masks.}
\label{fig:diversity_analysis}
\end{figure}

\textbf{6) Diversity validation through generation scale analysis:} The experimental results demonstrate that generation diversity reaches saturation around 1,500 to 2,000 masks, with performance gains exhibiting marginal returns beyond this range. We conducted diversity evaluation experiments by generating varying quantities of coronary artery masks (500, 1,000, 1,500, 2,000, and 2,500) and computed Inception Score (IS) to quantify mask diversity and quality. As shown in Fig.\ref{fig:diversity_analysis} (a), IS increased from 1.61±0.06 at 500 masks to 1.65±0.06 at 2,000 masks, with a slight decrease to 1.64±0.02 at 2,500 masks. This change indicates that when the number of generated masks exceeds 2,000, the vascular masks generated based on real vascular conditions exhibit repetitive structural patterns, leading to a decline in the upward trend of the IS value. For downstream segmentation evaluation, we used the synthesized masks to pre-train the generative model, fine-tuned this model using 5 coronary artery images, subsequently synthesized 500 masks to train a U-Net segmentation network, and evaluated the performance on 295 images from the SBCD test set. As shown in Fig. \ref{fig:diversity_analysis} (b), DSC increased from 82.23±4.53\% at 500 masks to 82.71±4.21\% at 2,000 masks and 82.74±4.36\% at 2,500 masks, with primary performance gains concentrated in the 1,500 to 2,000 mask range, while improvements exhibited diminishing returns between 2,000 and 2,500 masks. These results indicate that using approximately 1,500 to 2,000 synthesized coronary artery masks provides sufficient structural priors to effectively support generative learning and coronary artery segmentation on the SBCD dataset.

\section{CONCLUSIONS}
\label{sec:disscussion}
In this paper, we propose UniVG, a novel generative data-engine foundation model for universal few-shot vascular image segmentation. UniVG effectively addresses the challenges of vascular image segmentation in data-scarce scenarios by leveraging compositional learning and few-shot generative adaptation. By integrating generative modeling with large-scale vascular pre-training, our approach significantly enhances data diversity, model generalization, and segmentation accuracy with minimal annotated data. To support this, we develop UniVG-58K, a dataset comprising 57,075 unlabeled vascular images and a large collection of pseudo-vascular structures, providing the field with a comprehensive pre-training resource for future research. Extensive experiments on 11 vascular segmentation tasks validate the effectiveness of UniVG, demonstrating its ability to generate diverse vascular images and significantly improve few-shot segmentation performance. Notably, our method achieves results comparable to or exceeding fully supervised models using only five annotated images, highlighting its efficiency and practicality.

By providing a scalable, generalizable, and data-efficient solution for vascular segmentation, UniVG represents a promising advancement in the field. Future work will explore extensions to 3D vascular and further optimization of generative adaptation strategies to enhance domain-specific segmentation.

\section*{Acknowledgments}
This work was supported in part by the State Key Project of Research and Development Plan under Grants 2022YFC2401601; in part by the National Natural Science Foundation of China under Grant 62101249, and Grant T2225025; in part by the Natural Science Foundation of Jiangsu Province under Grant BK20210291; in part by the Jiangsu Provincial Double-Innovation Doctor Program under Grant JSSCBS20220202; in part by the Key Research and Development Programs in Jiangsu Province of China under Grant BE2021703 and BE2022768; in part by China Postdoctoral Science Foundation under Grant 2021TQ0149 and Grant 2022M721611, and by the Big Data Computing Center of Southeast University.

\appendix
\label{sec:experiments}
\begin{table*}[h!]
    \centering
    \caption{UniVG-58K Dataset and 11 Downstream Datasets: Covering vascular Images of Five Modalities. N/A entries marked with "\textemdash".}
    \resizebox{\linewidth}{!}
    {
    \begin{tabular}{l l c c c c c c}
    \toprule
    Dataset & Modality & Pre-training & Downstream task & Pre-training number & Downstream number & Fine-tune images & Test \\
    \midrule
    DiaRetDB1 \citep{kauppi2007diaretdb1} & Fundus & \checkmark & & 89 & \textemdash & \textemdash & \textemdash  \\
    e-Ophtha \citep{decenciere2013teleophta} & Fundus & \checkmark & & 463 & \textemdash & \textemdash & \textemdash  \\
    IDRID \citep{porwal2018indian} & Fundus & \checkmark & & 516  & \textemdash & \textemdash & \textemdash \\
    MESSIDOR \citep{decenciere2014feedback} & Fundus & \checkmark & & 1200 & \textemdash & \textemdash & \textemdash  \\
    Refuge \citep{orlando2020refuge} & Fundus  & \checkmark & &1200 & \textemdash & \textemdash & \textemdash  \\
    APTOS \citep{karthik2019aptos} & Fundus & \checkmark & & 3662  & \textemdash  & \textemdash & \textemdash\\
    Eyepacs \citep{gulshan2016development} & Fundus & \checkmark & & 35129 & \textemdash & \textemdash & \textemdash  \\
    DRIVE \citep{asad2014ant} & Fundus &  & \checkmark& \textemdash  &40 &5  &35  \\
    CHASEDB1 \citep{henry2020mixmodule} & Fundus&  & \checkmark & \textemdash  &28 &5 &23 \\
    HRF \citep{li1903connection} & Fundus&  & \checkmark & \textemdash  & 45 &5 &40 \\
    ORVS \citep{sarhan2021transfer} & Fundus &  & \checkmark & \textemdash  & 49 &5 &44 \\
    ARIA \citep{bankhead2012fast} & Fundus &  & \checkmark & \textemdash   & 142 &5 &137 \\
    XACD \citep{ma2021self} & Coronary Artery DSA& \checkmark & \checkmark & 1621 & 126 &5 &121 \\
    ARCADE \citep{maxim2023arcade} & Coronary Artery DSA & \checkmark & & 3000 & \textemdash   & \textemdash  & \textemdash\\
    SBCD & Coronary Artery DSA & &\checkmark  &  \textemdash & 300 &5 &295 \\
    ADSD \citep{Danilov2021AngiographicDF} & Coronary Artery DSA  & \checkmark &  & 8325 & \textemdash  & \textemdash & \textemdash \\
    DIAS \citep{liu2024dias} & Brain DSA & \checkmark & \checkmark & 441 & 60 &5 &55 \\
    DSCA \citep{xie2024dsnet, xie2024dsca} & Brain DSA &  & \checkmark & \textemdash & 224 &5 &219\\
    OCT \citep{li2024octa} & OCT & \checkmark & \checkmark & 600 & 300 &5 &295 \\
    OCTA500 \citep{li2024octa} &  OCTA & \checkmark &  \checkmark & 600 & 300 &5 &295  \\
    ROSE \citep{ma2020rose} &  OCTA &\checkmark & & 229 & \textemdash  & \textemdash  & \textemdash \\
    \midrule
    Total & & & & 57075 & 1614 & 55 & 1559\\
    \bottomrule
    \end{tabular}
    }
\label{tab:pre-vascular}
\end{table*}

\section{More details of UniVG datasets}
\label{appendix:More details of UniVG datasets}
We introduce a series of diverse and representative datasets that constitute UniVG-58K. These datasets cover five vascular modalities: fundus imaging, OCTA, OCT, brain DSA, and coronary artery DSA. Each dataset has been carefully preprocessed to ensure consistency in resolution and format. Below, we provide detailed descriptions of the datasets used for each imaging modality. 
\subsection{Fundus vascular}
We have integrated a series of publicly available and representative diabetic retinopathy datasets, which cover a wide range of imaging conditions and lesion characteristics. The selected datasets include DiaRetDB1, e-Ophtha, IDRID, MESSIDOR, Refuge, APTOS 2019, and Eyepacs, with detailed descriptions provided below:

DiaRetDB1  \citep{kauppi2007diaretdb1} is a publicly available benchmark dataset designed for diabetic retinopathy detection research. It contains 89 color fundus images (50° FOV) with resolutions of 1500×1152 pixels, collected from diverse clinical settings. Expert ophthalmologists have annotated lesions including microaneurysms, hemorrhages, and exudates, along with DR severity grading. 

e-Ophtha \citep{decenciere2013teleophta}  dataset, developed by the French Research Agency, is a publicly available ophthalmic image database for diabetic retinopathy (DR) and ocular lesion detection. It comprises two subsets:
e-Ophtha EX: 47 images with exudates (pathological cases) and 35 normal images. e-Ophtha MA: 148 images with microaneurysms (early DR signs) and 233 normal images All images were acquired with 45° FOV fundus cameras and annotated by expert ophthalmologists. The dataset provides pixel-level lesion masks for algorithm validation and is commonly used for automated DR screening studies. Access requires institutional agreement via the French Telemedicine Atlas platform.

IDRID  \citep{porwal2018indian} is a large-scale multimodal dataset for diabetic retinopathy and diabetic macular edema analysis. It contains 516 color fundus images (50° FOV) with ultra-high resolution (4288×2848 pixels), split into 413 training and 103 test images.

MESSIDOR dataset \citep{decenciere2014feedback} is a pioneering large-scale benchmark for diabetic retinopathy screening research. It contains 1,200 color fundus images acquired from three French medical centers using 45° FOV cameras with three resolutions: 1440×960, 2240×1488, and 2304×1536 pixels.

Refuge dataset \citep{karthik2019aptos}, from the MICCAI 2018/2020 challenges, is a multi-center benchmark for glaucoma detection and optic disc/cup segmentation. It contains 1,200 color fundus images with 45° FOV.

APTOS 2019 Blindness Detection dataset \citep{karthik2019aptos} is a large-scale competition dataset for diabetic retinopathy (DR) severity grading, organized by the Asia Pacific Tele-Ophthalmology Society. It contains 3,662 color fundus images collected from diverse clinical environments across India, with resolutions ranging from 1734×1734 to 5184×3456 pixels. Each image is graded by ophthalmologists into 5 DR severity levels (0: No DR to 4: Proliferative DR) following the International Clinical Diabetic Retinopathy Scale. 

Eyepacs \citep{gulshan2016development} dataset comprises 35,126 diabetic retinopathy (DR) fundus images, categorized into five severity levels (0-4). This open-source dataset was released through the 2019 Kaggle Diabetic Retinopathy Detection competition. Fundus images were captured by technicians from Aravind Eye Hospital, India, in rural areas with limited medical resources. Experienced ophthalmologists subsequently reviewed and annotated all images for DR grading. The dataset exhibits substantial variability in image resolution, ranging from 433 × 289 pixels to 5184 × 3456 pixels, reflecting real-world imaging conditions in resource-constrained environments.

DRIVE \citep{asad2014ant} dataset is a benchmark dataset for retinal vessel segmentation. It comprises a total of 40 color fundus images stored in JPEG format, including 7 pathological cases with abnormalities. These images were collected as part of a diabetic retinopathy screening program in the Netherlands. All images were captured using a Canon CR5 non-mydriatic 3CCD camera with a 45-degree field of view (FOV). Each image has a resolution of 584×565 pixels and 8-bit depth per color channel. The dataset is equally divided into a training set (20 images) and a test set (20 images).
Both sets include a circular FOV mask (approximately 540 pixels in diameter) for each image. In the training set, manual vessel segmentation was performed by a single ophthalmologist. For the test set, two independent observers manually segmented each image, with the first observer’s annotations serving as the gold standard for performance evaluation.

CHASEDB1 \citep{henry2020mixmodule} is a publicly available dataset for fundus image analysis, released by the Department of Computer Science at the University of Reading, UK. This dataset comprises 28 fundus images captured from both the left and right eyes of 28 subjects (including 14 healthy individuals and 14 patients with diabetic retinopathy). Each image has a resolution of 999×960 pixels. The primary objective of the CHASEDB1 dataset is to provide researchers with a standardized platform for developing and evaluating retinal vessel segmentation algorithms. Every fundus image in the dataset is accompanied by corresponding manually annotated vessel segmentation masks, which were meticulously labeled by qualified medical professionals.

HRF \citep{li1903connection} is a publicly available database specifically established for comparative studies of automated retinal image segmentation algorithms. The dataset currently comprises 15 images from healthy subjects, 15 images from patients with diabetic retinopathy, and 15 images from glaucoma patients. Each image is accompanied by a corresponding binary gold-standard vessel segmentation map. For specific subsets of the dataset, a field-of-view (FOV) mask is also provided to delineate the region of interest. All segmentation annotations were generated through collaboration between experts in retinal image analysis and clinicians from partner ophthalmology clinics, ensuring high reliability. Due to its rigorous annotation protocol and standardized format, HRF has become one of the most widely used benchmark datasets in the field of retinal vessel segmentation.

ORVS \citep{sarhan2021transfer}  dataset is a newly established collaborative effort between the Department of Computer Science and the Department of Vision Science at the University of Calgary. The dataset comprises 49 retinal images collected from a clinical facility in Calgary, Canada, with a division of 42 images for training and 7 images for testing. All images were acquired using a Zeiss Visucam 200 fundus camera with a 30-degree field of view (FOV). Each image has a resolution of 1444×1444 pixels and a 24-bit depth per pixel.
 
ARIA \citep{bankhead2012fast} dataset is a comprehensive retinal image collection comprising three distinct subsets: ARIA-AMD (23 cases with age-related macular degeneration), ARIA-DR (59 diabetic retinopathy cases), and ARIA-H (61 healthy controls). This multimodal dataset provides expert-annotated vascular segmentation masks and pathological labels, with images stored in uncompressed TIFF format at a resolution of 768×576 pixels.

\subsection{OCTA vascular}
ROSE \citep{ma2020rose}  dataset is an open-source OCTA retinal vessel segmentation dataset, comprising two subsets: ROSE-1 and ROSE-2. The ROSE-1 dataset consists of 117 OCTA images from 39 subjects (including 26 Alzheimer's disease patients and 13 healthy controls). Each subject has en-face (horizontal cross-sectional) angiograms of the superficial vascular complex (SVC), deep vascular complex (DVC), and an inner retinal slab combining both SVC and DVC. The scans cover a 3×3 mm² region centered on the fovea, with an image resolution of 304×304 pixels. ROSE-1 provides two types of vessel annotations: centerline-level and pixel-level annotations. The ROSE-2 dataset contains a total of 112 OCTA images. All images in this subset are en face angiograms of the SVC, acquired from a 3×3 mm² fovea-centered region. Only centerline-level annotations are provided.

OCTA-500 \citep{li2024octa} dataset comprises paired three-dimensional OCT and OCTA imaging data from 500 eyes, including six types of projection images, four categories of textual metadata (age, gender, eye laterality, and disease classification), and two types of segmentation labels (retinal large vessels and foveal avascular zone). Six distinct projection types were generated from the OCT and OCTA datasets by applying either average or maximum intensity projection methods across three anatomical boundaries: the inner limiting membrane, the outer plexiform layer, and Bruch's membrane. 

\subsection{OCT vascular}
OCT dataset from the OCTA-500 dataset \citep{li2024octa}  comprises 300 annotated three-dimensional scans with corresponding vascular segmentation labels. The dataset provides co-registered structural OCT and angiographic OCTA data, enabling simultaneous analysis of retinal morphology and vasculature.  

\subsection{Brain DSA vascular}
DIAS \citep{liu2024dias} is a dataset that contains 60 brain DSA mask annotations, each corresponding to five arterial phase frames (we manually selected vascular images that align well with the mask annotations; the original images have a resolution of 800×800 pixels, and both images and masks were resized to 512×512 pixels), as well as 441 unlabeled brain DSA images.

DSCA \citep{xie2024dsnet}, which contains 224 brain DSA images and their corresponding masks with varying resolutions, including 512×512, 1024×1024, 952×952, 512×472, 1232×1232, 1432×1432, and 742×960 pixels (all images and masks are resized to 512×512 pixels). For the few-shot evaluation, we randomly select 5 annotated images as the training set and use the remaining images as the test set. 

\subsection{Coronary artery DSA vascular dataset}
ARCADE dataset contains 3,000 X-ray coronary angiography images, designed for two tasks: coronary artery segmentation (1,500 images) with annotations of vascular branches according to the Syntax Score criteria, and stenosis detection (1,500 images) with pixel-level localization of plaques and stenotic lesions. The data was acquired using standard angiography techniques, with a resolution of 1024×1024 pixels.

XACD \citep{ma2021self} dataset contains coronary angiography images acquired during stent placement using the General Electric Innova IGS 520 system. Each image has a resolution of 512×512 pixels and is single-channel. This dataset includes 1,621 coronary angiography images and 126 independent coronary angiography images, accompanied by vessel segmentation maps annotated by experienced radiologists.

SBCD dataset, which originates from the Second People’s Hospital of Shenzhen and contains 300 X-ray angiography images (each image comes with corresponding expert segmentation annotations, and the dataset includes segmentation labels for both the left coronary artery (LCA) and the right coronary artery (RCA), with 150 images featuring the LCA and the other 150 images featuring the RCA; all images have a resolution of 512×512 pixels). For the few-shot evaluation, we randomly select 5 annotated images as the training set and use the remaining images as the test set.

\bibliographystyle{model2-names.bst}\biboptions{authoryear}

\bibliography{references}

\end{document}